\documentclass[10pt,letterpaper]{article}
\usepackage[top=0.85in,left=2.75in,footskip=0.75in]{geometry}
\usepackage{graphicx}
\usepackage{multirow}
\usepackage{subcaption}
\usepackage{booktabs}
\usepackage{ragged2e}
\usepackage{amsmath,amssymb}
\usepackage{diagbox}
\usepackage{changepage}
\usepackage{array}
\usepackage{textcomp,marvosym}

\usepackage[nospace]{cite}

\usepackage{nameref,hyperref}

\usepackage[nopatch=eqnum]{microtype}
\DisableLigatures[f]{encoding = *, family = * }

\usepackage[table]{xcolor}

\usepackage{array}

\newcolumntype{+}{!{\vrule width 2pt}}

\newlength\savedwidth

\raggedright
\usepackage[aboveskip=1pt,labelfont=bf,labelsep=period,justification=raggedright,singlelinecheck=off]{caption}

\makeatletter
\renewcommand{\@biblabel}[1]{\quad#1.}
\makeatother

\usepackage{lastpage,fancyhdr,graphicx}
\usepackage{epstopdf}
\fancyheadoffset[L]{2.25in}
\fancyfootoffset[L]{2.25in}


\begin{document}
\vspace*{0.2in}

\begin{flushleft}
{\Large
\textbf{Adaptive attention routing multimodal expert graph neural network for drug--drug interaction prediction}
}

\vspace{0.3cm}

Jingyi Yang\textsuperscript{1,\Yinyang},
Chengcheng Yan\textsuperscript{2,\Yinyang},
Hongri Cong\textsuperscript{1},
Feifei Zhao\textsuperscript{1,3*}
\\
\bigskip
\textbf{1} College of Science, Shihezi University, Shihezi, 832003, China
\\
\textbf{2} College of Artificial Intelligence, Shaoxing Institute of Technology, Shaoxing, 312000, China
\\
\textbf{3} School of Mathematics and Computational Science, Xiangtan University, Xiangtan, 411105, China
\\
\bigskip
\Yinyang These authors contributed equally to this work.
\\
* zhaofeifei@shzu.edu.cn
\end{flushleft}

\section*{Abstract}
\begin{justify}
Accurate prediction of drug–drug interactions (DDIs) is essential for drug safety in combination therapy. Multimodal methods built on knowledge graphs (KGs) already combine molecular structures, substructures, targets and biomedical relations into drug representations, but they still face three problems: features are heterogeneous across modalities, fusion does not adapt to individual drug pairs, and modality-specific information is diluted during integration. We propose the adaptive attention routing multimodal expert graph neural network (AARME-GNN) to address them. A Feature Adaptive Calibration Module (FACM) reduces within-modality noise in each representation through bottleneck reparameterization and channelwise recalibration before fusion. A Multimodal Weight Router (MMWR) then sets the modality weights of each drug pair, with an entropy constraint that prevents the weights from collapsing onto one modality. A Routing-Aware Weighted Distillation (RAWD) mechanism builds a soft target from the unimodal branches, so that the fusion classifier retains the knowledge held by each modality. On Dataset 1, AARME-GNN improves precision over MKG-FENN by 3.71 and 6.57 percentage points in known drug and known–new drug prediction, and the area under the receiver operating characteristic curve (AUC) by 4.53 in new drug prediction. On Dataset 2, the advantage is concentrated in the two cold-start settings and in precision.
\end{justify}

\section*{Author summary}
\begin{justify}
Combination therapy improves outcomes in many chronic diseases, but every drug added raises the risk of DDIs. Computational methods now predict DDI events from molecular and biomedical data, yet three problems remain. Heterogeneous modalities are difficult to calibrate, one fusion strategy serves every drug pair, and modality-specific information is diluted during fusion. We propose AARME-GNN, an adaptive attention routing multimodal expert graph neural network, to address all three. A calibration module reduces within-modality noise in each representation, a router sets modality weights for each drug pair, and a distillation term keeps the unimodal branches informative for the fusion classifier. We test the model on two public DDI benchmarks under three prediction settings that involve known drugs, known–new drug pairs and new drugs. On both benchmarks it stays close to the strongest baselines on most metrics and improves precision, at a modest cost in recall. The same components apply to other multimodal biomedical prediction problems, where modalities differ in reliability from one sample to the next.
\end{justify}
\section*{Introduction}
\begin{justify}
\label{sec1}
Combination therapy is widely used for patients with complex diseases, chronic comorbidities, or polypharmacy, but it increases the risk of DDIs \cite{nicholson2024prevalence,guthrie2015rising,mallet2007challenge}. DDIs may alter drug absorption, distribution, metabolism, and excretion, as well as pharmacodynamics, leading to reduced efficacy, increased toxicity, or other adverse clinical outcomes \cite{juurlink2003drug,becker2007hospitalisations}. Clinically relevant DDIs have been associated with hospitalizations related to adverse drug reactions, especially in elderly populations \cite{hughes2024drug,cosgrave2025hospital}. Therefore, accurate and scalable DDI prediction is essential for improving drug safety in combination therapy.

Traditional experimental approaches are reliable but costly and limited in coverage. With the growth of biomedical KGs and multimodal drug representations, learning-based methods have become the practical route to large-scale DDI prediction. Multimodal integration enables richer representation learning by combining molecular structures, substructures, targets, enzymes, pathways, and biomedical relations \cite{lin2023comprehensive}.

Early methods relied on handcrafted features and shallow models, limiting their ability to capture complex topological and semantic dependencies in heterogeneous drug data \cite{lin2023comprehensive,ryu2018deep,deng2020multimodal}. Graph neural networks (GNNs) have become the standard way to model relational structure. Decagon models polypharmacy side effects on a multi-relational graph \cite{zitnik2018modeling}. KGNN learns drug representations via neighborhood aggregation over biomedical entities \cite{lin2020kgnn}. SumGNN improves prediction by extracting informative subgraphs \cite{yu2021sumgnn}. More recently, multimodal KG methods combine molecular graphs with KGs for enhanced representation learning. KGMAEDDI, for instance, uses masked autoencoding to encode molecular and knowledge graphs jointly \cite{li2025kgmaeddi}. TIGER applies a relation-aware heterogeneous graph Transformer to DDI prediction \cite{su2024dual}, while MKG-FENN \cite{wu2024mkg} fuses multiple modality-specific graphs and MeTDDI \cite{zhong2024learning} learns motif-based graphs for interpretable DDI prediction.

\begin{figure}[!ht]
\begin{adjustwidth}{-5.9cm}{0cm}
\centering
\captionsetup{width=1.4\textwidth, justification=justified}
\includegraphics[height=2.754in]{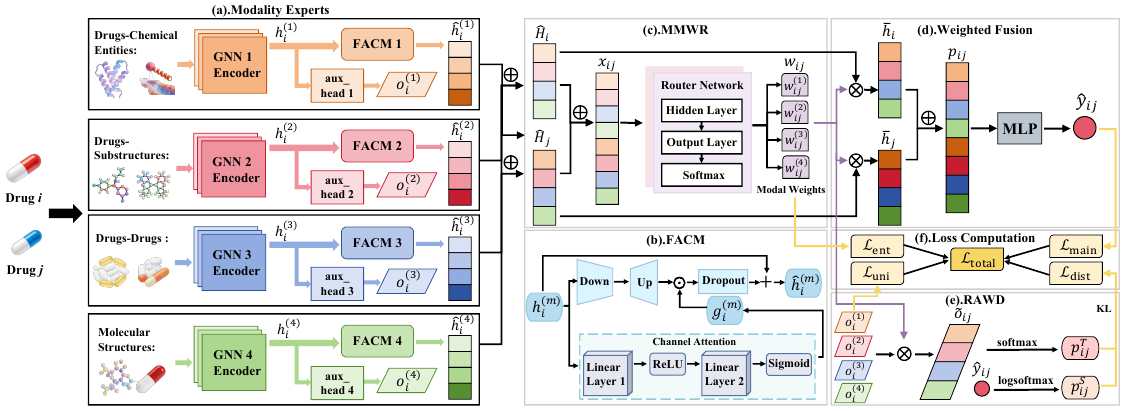}\vspace*{1pt}
\caption{{\bf Overall architecture of AARME-GNN.}
(a) Modality experts extract modality-specific drug representations from multi-source KGs.
(b) FACM calibrates each modality representation before fusion to reduce within-modality noise and enhance informative channel responses.
(c) MMWR generates drug pair-specific modality weights from calibrated representations.
(d) Weighted fusion aggregates modality-specific representations into a drug pair-level multimodal representation.
(e) RAWD combines the unimodal branches into a routing-weighted soft target $\textit{p}_{ij}^{T}$, to which the prediction distribution $\textit{p}_{ij}^{S}$ of the main fusion classifier is matched.
(f) The training objective jointly optimizes the main classification loss, unimodal auxiliary loss, routing-aware distillation loss, and routing entropy regularization term.}
\label{fig:AARME}
\end{adjustwidth}
\end{figure}

Despite these advances, two core challenges remain. First, the modalities of a drug differ in distribution and in meaning, which makes their features heterogeneous and degrades fusion \cite{lin2023comprehensive,li2024multimodal,planamente2024relative}. Second, most methods rely on fixed or globally shared fusion strategies, which cannot adaptively model the importance of each modality for individual samples and may suppress complementary modality information \cite{wu2024mkg,shen2025benchmarking}. The first challenge has two distinct aspects, which we address separately. Within a modality, noisy channels and unstable feature scaling are handled at the representation level by FACM, which calibrates each modality independently and uses neither a cross-modal alignment objective nor a shared projection layer. Across modalities, the relevance of each modality varies from one drug pair to the next. This is handled at the fusion level by MMWR, which re-balances the modality contributions of each drug pair instead of mapping all modalities into a common space.

Recent studies show that lightweight adaptation and feature calibration improve multimodal representation learning \cite{wu2025medical,he2025adaptfrcnet}, and that channelwise recalibration helps select informative features and suppress noise \cite{hu2018squeeze,planamente2024relative}. Motivated by these findings, we propose AARME-GNN, an adaptive multimodal framework for DDI prediction. FACM calibrates each modality representation before fusion. MMWR assigns modality weights tailored to each drug pair for adaptive fusion. RAWD transfers unimodal knowledge to the fusion model via routing-aware distillation. By integrating these components, AARME-GNN improves feature calibration, fusion adaptability, and knowledge preservation across modalities. The contributions of this paper are as follows:

\begin{itemize}
    \item We propose AARME-GNN for multi-class DDI event prediction. The model addresses three problems of multimodal KG-based prediction: noisy and unevenly scaled unimodal representations, fusion that does not adapt to individual drug pairs, and the loss of modality-specific information during integration.
    \item We design three modules that work together. FACM calibrates each modality representation before fusion, MMWR assigns modality weights to each drug pair, and RAWD transfers discriminative knowledge from the unimodal branches to the fusion classifier.
    \item On two public DDI benchmarks and three prediction settings, AARME-GNN attains the highest AUC and the highest precision in every setting of Dataset~1, the highest accuracy in two of its three settings, and the highest accuracy on the new drug setting of Dataset~2.
\end{itemize}
\end{justify}

\section*{Related Work}

\label{sec2}

\subsection*{Traditional Deep Learning for DDI Prediction}
\begin{justify}

Early studies on DDI prediction mainly relied on drug similarities and handcrafted molecular features. Vilar et al. developed a scalable similarity-based framework by integrating two-dimensional and three-dimensional molecular structures, interaction profiles, target similarity, and adverse-effect similarity from known interaction databases \cite{vilar2014similarity}. Although such approaches provide interpretability, their performance is limited by feature engineering and incomplete coverage of known interaction data.

Deep learning replaced handcrafted descriptors with learned drug representations. DeepDDI applies drug structural information to deep neural networks for multi-class DDI and drug--food interaction prediction \cite{ryu2018deep}, demonstrating that neural models can automatically learn interaction patterns from molecular representations. DDIMDL further integrates chemical substructures, targets, enzymes, and pathways to construct multimodal drug representations for DDI prediction \cite{deng2020multimodal}. These approaches mark a transition from manual feature engineering to data-driven representation learning.

Attention-based and few-shot methods were then introduced to improve robustness and the prediction of rare interactions. MDF-SA-DDI employs multi-source feature fusion with Transformer-based self-attention for DDI event prediction \cite{lin2022mdf}, while META-DDIE introduces a comparison-based mechanism to improve rare DDI event recognition under limited labeled data \cite{deng2022meta}. Despite these advances, most methods still rely on fixed fusion structures and implicit interaction modeling, limiting their ability to capture sample-specific modality importance and high-order relational dependencies.

\end{justify}

\subsection*{Graph-based Representation Learning for DDI Prediction}
\begin{justify}

Graph-based methods now dominate DDI prediction because they model the relational structure of biomedical data directly. Decagon constructs a multi-relational graph integrating drug--drug, drug--target, and protein--protein interactions for polypharmacy side-effect prediction \cite{zitnik2018modeling}. KGNN learns drug representations by aggregating neighborhood information from biomedical KGs \cite{lin2020kgnn}, while SumGNN improves reasoning capability through knowledge subgraph extraction and summarization \cite{yu2021sumgnn}.

To capture fine-grained molecular interaction patterns, SSI-DDI directly models substructure-level interactions from raw molecular graphs without relying on handcrafted descriptors \cite{nyamabo2021ssi}. Molormer introduces self-attention-based modeling of molecular structures \cite{zhang2022molormer}. DDKG and LaGAT further incorporate biomedical KGs and drug pair-specific attention mechanisms to improve semantic reasoning and interpretability \cite{su2022attention,hong2022lagat}.

DGNN-DDI \cite{ma2023dual} introduces a dual graph neural network that models molecular structure and substructure interactions through directed message passing with substructure attention. A co-attention mechanism is further used to capture interaction patterns between drug substructures, showing that explicit substructure-level interaction modeling can improve both prediction accuracy and interpretability.

More recent methods extend GNNs to heterogeneous biomedical settings. MRCGNN integrates molecular graphs with multi-relational DDI graphs through contrastive learning \cite{xiong2023multi}, while TIGER adopts relation-aware heterogeneous graph Transformers to capture long-range dependencies in biomedical graphs \cite{su2024dual}. MeTDDI introduces motif-based structural learning and co-attention mechanisms for interpretable DDI prediction \cite{zhong2024learning}. However, existing methods still suffer from limited robustness under distribution shifts and insufficient modeling of modality heterogeneity \cite{shen2025benchmarking}.

\end{justify}

\subsection*{Multimodal and Heterogeneous Graph Learning for DDI Prediction}
\begin{justify}

To incorporate richer biomedical context, heterogeneous and multimodal graph learning methods have been widely explored. MUFFIN, MDNN, and MDF-SA-DDI integrate molecular graphs, drug attributes, and biomedical knowledge to learn multimodal drug representations for DDI prediction \cite{chen2021muffin,lyu2021mdnn,lin2022mdf}. MKG-FENN constructs multiple modality-specific graphs, including drug--chemical entity, drug--substructure, DDI, and molecular structure graphs, and performs multi-channel graph learning for feature fusion \cite{wu2024mkg}. TIGER further integrates molecular graphs and biomedical KGs using heterogeneous Transformer architectures \cite{su2024dual}.

HBGCN \cite{zhu2026heterogeneous} constructs a heterogeneous biological network consisting of drugs, genes, proteins, and diseases, and introduces a hierarchical graph convolutional framework to model both direct and meta-path-level dependencies. Heterogeneous representation learning therefore carries over from interaction modeling to the more structured biomedical setting.

Despite these advances, most existing methods rely on static fusion strategies or implicit attention mechanisms, which limit their ability to model sample-specific modality contributions. KA-DDI refines the knowledge graph adaptively through contrastive learning to suppress task-irrelevant relations, but without explicit sample-level modality routing \cite{li2026ka}. Channelwise recalibration re-weights informative channels and improves feature quality \cite{hu2018squeeze}, which motivates feature calibration in the multimodal setting.

Beyond DDI prediction, EMOE introduces sample-level modality routing and unimodal distillation strategies to balance modality contributions while preserving modality-specific knowledge \cite{fang2025emoe}. EMOE was designed for general multimodal learning, but its two mechanisms map directly onto the fusion problem of DDI prediction.

Existing studies still lack a single framework that addresses feature-space calibration, sample-level modality routing and modality-specific knowledge preservation together. This limitation motivates the proposed AARME-GNN, which integrates adaptive calibration, routing, and distillation mechanisms to improve multimodal representation learning for DDI prediction.

\end{justify}

\section*{Materials and methods}

\subsection*{Overview}
\begin{justify}

This section presents AARME-GNN for DDI event prediction. As shown in Fig~\ref{fig:AARME}, the model is built on four heterogeneous KGs and aims to improve multimodal feature calibration, sample-level fusion adaptiveness, and modality-specific knowledge preservation. The overall framework consists of modality-specific graph encoding, FACM-based feature calibration, MMWR-based modality routing, weighted fusion and classification, and RAWD-based training regularization.

Let $\mathcal{D}$ denote the drug set and $N_{d}$ be the number of drug nodes. For a candidate drug pair $(d_i, d_j)$, modality experts first extract drug representations from four heterogeneous KGs. FACM then calibrates each modality representation before fusion to reduce within-modality noise. Based on the calibrated representations, MMWR generates drug pair-specific modality weights and performs dynamic weighted fusion. The main fusion classifier predicts the DDI event class from the fused drug pair representation. During training, RAWD aggregates the unimodal auxiliary predictions into a routing-weighted soft target and transfers this discriminative unimodal knowledge to the main fusion classifier. Thus, AARME-GNN forms a unified pipeline that integrates encoding, calibration, routing, fusion-based classification, and distillation-based regularization.

For consistent notation, the modality index is denoted as $m \in \{1,2,3,4\}$. The raw representation matrix of all drugs under the $m$-th modality is defined as
\begin{equation}
\mathbf{H}^{(m)} =
\left[
\textit{h}_{1}^{(m)}, \textit{h}_{2}^{(m)}, \ldots, \textit{h}_{N_{d}}^{(m)}
\right]^{\top} \in \mathbb{R}^{N_{d} \times d_{e}},
\end{equation}
where $\textit{h}_i^{(m)}$ is the raw representation of drug $d_i$ under the $m$-th modality. Throughout this work, the embedding dimensionality is set to $d_{e}=128$, and the model outputs a predicted distribution over $C=65$ DDI event classes for each drug pair $(d_i, d_j)$.

\end{justify}

\subsection*{Feature Adaptive Calibration Module}
\begin{justify}

Drug representations learned from different knowledge graphs vary in distribution, semantic granularity and discriminative preference, and individual modalities may contain noisy or weakly scaled channels. FACM adapts every modality representation before fusion through lightweight feature mapping and channelwise calibration. Inspired by feature calibration in vision tasks \cite{he2025adaptfrcnet}, we extend this idea to multi-KG representation learning: FACM first maps each modality representation into a compact latent space to improve feature adaptability, and then recalibrates its channels. This process reduces within-modality noise and provides better-scaled inputs for the subsequent routing and fusion.

Following the notation established in Eq.~(1), $\mathbf{H}^{(m)}$ denotes the raw representation matrix of the $m$-th modality, where the $i$-th row $\textit{h}_i^{(m)}$ corresponds to the raw feature vector of drug $d_i$. FACM takes each row vector $\textit{h}_i^{(m)}$ as input and performs adaptive calibration to obtain a modality-calibrated representation. Specifically, for a given drug $d_i$, the raw representation $\textit{h}_i^{(m)}$ is first mapped into a low-dimensional latent space:
\begin{equation}
\textit{z}_i^{(m)} =
\operatorname{Dropout} \Big(
\delta \big( \mathbf{W}_d^{(m)} \textit{h}_i^{(m)} + \textit{b}_d^{(m)} \big)
\Big),
\end{equation}
where $\mathbf{W}_{d}^{(m)} \in \mathbb{R}^{r \times d_{e}}$ and $\textit{b}_d^{(m)} \in \mathbb{R}^{r}$ are the weight matrix and bias vector of the down-projection mapping, respectively. Here, $r$ denotes the bottleneck dimension, $\delta(\cdot)$ is the ReLU activation function, $\operatorname{Dropout}(\cdot)$ denotes the dropout operation, and $\textit{z}_i^{(m)} \in \mathbb{R}^{r}$ is the resulting latent representation. In this implementation, $r=64$.

Subsequently, the latent representation $\textit{z}_i^{(m)}$ is projected back to the original feature space to obtain the reconstructed feature $\textit{u}_i^{(m)} \in \mathbb{R}^{d_{e}}$:
\begin{equation}
\textit{u}_i^{(m)} = \mathbf{W}_u^{(m)} \textit{z}_i^{(m)} + \textit{b}_u^{(m)},
\end{equation}
where $\mathbf{W}_{u}^{(m)} \in \mathbb{R}^{d_{e} \times r}$ and $\textit{b}_u^{(m)} \in \mathbb{R}^{d_{e}}$ are the weight matrix and bias vector of the up-projection mapping, respectively. FACM then uses channel attention to estimate the importance of each feature dimension:
\begin{equation}
\textit{g}_i^{(m)} =
\sigma \Big(
\mathbf{W}_2^{(m)}
\delta ( \mathbf{W}_1^{(m)} \textit{h}_i^{(m)} + \textit{b}_1^{(m)} )
+ \textit{b}_2^{(m)}
\Big),
\end{equation}
where $\mathbf{W}_{1}^{(m)} \in \mathbb{R}^{s \times d_{e}}$, $\textit{b}_1^{(m)} \in \mathbb{R}^{s}$, $\mathbf{W}_{2}^{(m)} \in \mathbb{R}^{d_{e} \times s}$, and $\textit{b}_2^{(m)} \in \mathbb{R}^{d_{e}}$ are the parameters of the channel-attention network. Here, $s$ is the hidden-layer dimension, $\sigma(\cdot)$ denotes the Sigmoid function, and $\textit{g}_i^{(m)} \in \mathbb{R}^{d_{e}}$ represents the channel-weight vector. In this implementation, $s=\max(4,d_{e}/8)$, which gives $s=16$ when $d_{e}=128$.

The reconstructed feature is then modulated by the channel-weight vector and added to the original input through a residual connection:
\begin{equation}
\widehat{\textit{h}}_i^{(m)} =
\textit{h}_i^{(m)} + \operatorname{Dropout} \big( \textit{u}_i^{(m)} \odot \textit{g}_i^{(m)} \big),
\end{equation}
where $\odot$ denotes element-wise multiplication. This design calibrates modality-specific representations without changing the original graph encoder. The channel-attention vector is deliberately computed from the raw representation rather than from the reconstruction. Its role is to decide which channels of the input require correction, and the modulated reconstruction is added back to that same input through the residual connection of Eq.~(5). Computing the gate from $\textit{u}_i^{(m)}$ would make it self-referential, since the attention weights could then be explained by the very reconstruction they modulate, which would weaken the intended denoising effect. Evaluating the gate on $\textit{h}_i^{(m)}$ additionally retains information that the low-rank bottleneck may discard. By combining bottleneck reparameterization, channel recalibration and residual learning, FACM preserves the original topological and semantic information while providing more stable and comparable inputs for subsequent dynamic routing. FACM operates independently on each modality, with its own parameters, no cross-modal alignment objective and no shared projection layer. Its role is to improve the quality of individual modality representations before routing and fusion, not to map different modalities into a common space.
\end{justify}

\subsection*{Multimodal Weight Router}
\begin{justify}

Improving the compatibility of modality representations alone is insufficient to capture sample heterogeneity in DDI prediction. Different DDI events may rely more on molecular structures, targets, pathways, or drug-relation semantics. Therefore, after feature calibration, the model needs to estimate the contribution of each modality according to the current drug pair. MMWR is designed for this purpose. It generates a sample-level modality-weight distribution from FACM-calibrated representations and performs dynamic weighted fusion.

For a drug pair $(d_i, d_j)$, the calibrated modality-specific representations are denoted as $\widehat{\textit{h}}_i^{(m)}$ and $\widehat{\textit{h}}_j^{(m)}$, respectively. The representations of the two drugs from the four modalities are concatenated in a fixed order to obtain $\widehat{\textit{H}}_i$ and $\widehat{\textit{H}}_j$, respectively, which together form the routing input $\textit{x}_{ij} \in \mathbb{R}^{8d_{e}}$.

A two-layer feed-forward network is then used to generate raw routing scores for the four modalities:
\begin{equation}
\textit{s}_{ij} =
\mathbf{W}_{r2} \delta ( \mathbf{W}_{r1} \textit{x}_{ij} + \textit{b}_{r1} ) + \textit{b}_{r2},
\end{equation}
where $\mathbf{W}_{r1} \in \mathbb{R}^{d_{r} \times 8d_{e}}$ and $\textit{b}_{r1} \in \mathbb{R}^{d_{r}}$ are the parameters of the first layer, while $\mathbf{W}_{r2} \in \mathbb{R}^{4 \times d_{r}}$ and $\textit{b}_{r2} \in \mathbb{R}^{4}$ are the parameters of the second layer. Here, $d_{r}$ is the hidden-layer dimension of the router, and $\textit{s}_{ij} \in \mathbb{R}^{4}$ denotes the raw modality scores. In this implementation, $d_{r}=64$.

Softmax normalization is applied to $\textit{s}_{ij}$ to obtain the sample-level modality-weight distribution $\textit{w}_{ij} \in \mathbb{R}^{4}$. Each weight $w_{ij}^{(m)}$ represents the relative contribution of the $m$-th modality to the prediction of drug pair $(d_i, d_j)$, and all modality weights sum to one.

After obtaining the modality weights, the same set of drug pair-specific weights $\textit{w}_{ij}$ is applied to both drugs. The dynamic fused representations are computed as:
\begin{equation}
\overline{\textit{h}}_i = \sum_{m=1}^{4} w_{ij}^{(m)} \cdot
\widehat{\textit{h}}_i^{(m)}, \quad
\overline{\textit{h}}_j = \sum_{m=1}^{4} w_{ij}^{(m)} \cdot
\widehat{\textit{h}}_j^{(m)},
\end{equation}
where the shared weights $w_{ij}^{(m)}$ ensure that both drugs' representations are fused under consistent modality evidence.

These two representations are concatenated to form the drug pair-level fused representation:
\begin{equation}
\textit{p}_{ij} = \overline{\textit{h}}_i \oplus \overline{\textit{h}}_j \in \mathbb{R}^{2d_{e}},
\end{equation}
where $\oplus$ denotes vector concatenation.

The fused representation $\textit{p}_{ij}$ is then fed into the main fusion classifier, implemented as a multi-layer perceptron (MLP), to obtain the main classification logits $\textit{o}_{ij}^{M} \in \mathbb{R}^{C}$. The predicted probability distribution of drug pair $(d_i,d_j)$ is computed as
\begin{equation}
\hat{\textit{y}}_{ij} = \operatorname{Softmax}(\textit{o}_{ij}^{M}) \in \mathbb{R}^{C}.
\end{equation}

To prevent the router from prematurely assigning most weights to a single modality, an entropy regularization term is introduced to constrain the modality-weight distribution during training. Let $\mathcal{B}$ denote a training batch, $|\mathcal{B}|$ denote the batch size, and $\varepsilon$ denote a numerical stability constant. The routing entropy regularization loss is defined as
\begin{equation}
\mathcal{L}_{\mathrm{ent}}
=
\frac{1}{|\mathcal{B}|}
\sum_{(i,j) \in \mathcal{B}}
\sum_{m=1}^{4}
w_{ij}^{(m)}
\log
\left(
w_{ij}^{(m)}
+
\varepsilon
\right).
\end{equation}

This formulation corresponds to the negative entropy of the modality-weight distribution. Minimizing $\mathcal{L}_{\mathrm{ent}}$ drives the modality weights toward a uniform distribution ($w_{ij}^{(m)} \to 1/4$), counteracting the tendency of gradient-based optimization to concentrate weight on a single modality. The coefficient $\lambda_{\mathrm{balance}}$ controls the strength of this regularization, with larger values yielding more balanced routing. In this implementation, $\varepsilon=10^{-8}$.

The routing input $\textit{x}_{ij}$ concatenates the calibrated representations of $d_i$ and $d_j$ in a fixed order, with the four modalities of $d_i$ followed by those of $d_j$. The reversed pair $(d_j, d_i)$ therefore yields a different routing input and a potentially different weight distribution, while the two drugs of each ordered pair share the same weights. The convention that handles the two directions of an undirected DDI event during training and inference is specified at the end of the \emph{Dataset} subsection.
\end{justify}

\subsection*{Routing-Aware Weighted Distillation}
\begin{justify}

With MMWR, the model can assign drug pair-specific modality weights and generate sample-level fused representations. However, dynamic fusion mainly operates at the feature level and may not fully preserve discriminative knowledge from each unimodal branch. In DDI prediction, different modalities provide complementary cues from molecular structures, substructures, drug relations, and biomedical semantics. During fusion, modality-specific cues that are useful for certain DDI events may be compressed or weakened. If the model is supervised only by the main fusion classifier, it may emphasize the global discriminability of the fused representation while underusing informative unimodal knowledge.

To address this issue, RAWD is introduced on top of MMWR-based dynamic routing and reuses the sample-level modality weights to guide knowledge transfer. It combines the auxiliary predictions of the four unimodal branches into a routing-weighted soft target and trains the main fusion classifier to match this target through a distillation constraint. In this way, RAWD encourages the fused representation to retain useful unimodal information while maintaining the adaptive fusion behavior learned by MMWR.

Each modality encoder is equipped with an auxiliary prediction head for drug-level supervision. For an ordered drug pair $(d_i, d_j)$, with $d_i$ the first drug in the current order, the unimodal auxiliary logits of modality $m$ are computed as
\begin{equation}
\textit{o}_i^{(m)} = \mathbf{W}_a^{(m)} \textit{h}_i^{(m)} + \textit{b}_a^{(m)},
\end{equation}
where $\mathbf{W}_{a}^{(m)} \in \mathbb{R}^{C \times d_{e}}$ and $\textit{b}_a^{(m)} \in \mathbb{R}^{C}$ are the parameters of the $m$-th auxiliary head and $\textit{o}_i^{(m)} \in \mathbb{R}^{C}$ is its prediction for the first drug. Because both orders of every training pair are used, each drug is supervised under different pairwise contexts, which enables the auxiliary heads to learn drug-level features related to DDI events. The head receives only the representation of $d_i$ but is trained on the pair label $y_{ij}$. Because the same drug appears with different partners, the supervision is potentially conflicting, and the head learns features that remain discriminative across pairwise contexts while also regularizing the fusion network.

The modality weights generated by MMWR are then used to aggregate the four unimodal auxiliary outputs into a sample-adaptive soft target:
\begin{equation}
\widetilde{\textit{o}}_{ij} = \sum_{m=1}^4 w_{ij}^{(m)} \textit{o}_i^{(m)},
\end{equation}
where $\widetilde{\textit{o}}_{ij} \in \mathbb{R}^{C}$ denotes the target logits for the current sample. Because the routing weights are computed from the drug pair, the target reflects the modality-specific discriminative preference of that pair and, through the weights, depends on both drugs rather than on $d_i$ alone.

RAWD is therefore a self-distillation objective: the target is produced by the model's own auxiliary heads, which share the modality encoders with the main classifier, and no separate network is trained for it. The only additional parameters are the four lightweight auxiliary heads, which are trained by the unimodal auxiliary loss $\mathcal{L}_{\mathrm{uni}}$ rather than by the distillation loss. The target distribution is detached from the computational graph, so the gradients of $\mathcal{L}_{\mathrm{dist}}$ do not flow into the auxiliary branches and the main classifier is matched against a fixed target at each step. The knowledge transferred is thus the relative, soft class preference of the routing-weighted unimodal ensemble for the current drug pair, a response-based distillation signal in the taxonomy of knowledge distillation \cite{hinton2015distilling,gou2021knowledge}, and not the decision boundary of a separate model. In this sense RAWD belongs to the family of self-distillation methods \cite{furlanello2018born,zhang2019be}, with the difference that the target is not a previous version of the network but the routing-weighted ensemble of its unimodal branches, and that the weighting is sample-dependent.

The target logits and the logits of the main fusion classifier are then normalized to obtain the target distribution $\textit{p}_{ij}^{T} \in \mathbb{R}^{C}$ and the prediction distribution $\textit{p}_{ij}^{S} \in \mathbb{R}^{C}$ of the main classifier, respectively. In the implementation, the main output is used in its log-probability form for the Kullback--Leibler (KL) divergence computation. The distillation loss is defined as
\begin{equation}
\mathcal{L}_{\mathrm{dist}}
=
\frac{1}{|\mathcal{B}|}
\sum_{(i,j) \in \mathcal{B}}
\mathrm{KL}
\left(
\textit{p}_{ij}^{T}
\|
\textit{p}_{ij}^{S}
\right),
\end{equation}
where $\operatorname{KL}(\cdot \| \cdot)$ denotes the Kullback--Leibler divergence. Through this constraint, the routing-weighted soft target provides additional supervision for the main fusion classifier.

To preserve the discriminative capability of each modality branch, unimodal supervision is also imposed on the four auxiliary prediction heads. Let $y_{ij}$ denote the ground-truth label of drug pair $(d_i,d_j)$. The unimodal auxiliary loss is defined as
\begin{equation}
\mathcal{L}_{\mathrm{uni}} = \frac{1}{4} \sum_{m=1}^{4}\frac{1}{|\mathcal{B}|} \sum_{(i,j) \in \mathcal{B}} \mathrm{CE}(\textit{o}_i^{(m)}, y_{ij}),
\end{equation}
where $\mathrm{CE}(\cdot,\cdot)$ denotes the cross-entropy loss function.

During inference, only the four modality encoders, FACM, MMWR, and the main fusion classifier are retained. For a candidate drug pair $(d_i,d_j)$, the model extracts four modality-specific representations, calibrates them with FACM, and dynamically fuses them using MMWR. The main fusion MLP then outputs the predictive probability distribution and determines the DDI event class. Therefore, AARME-GNN preserves the original multi-graph encoding framework while transforming static fusion into a collaborative paradigm that integrates calibration, routing, and distillation.
\end{justify}

\subsection*{Loss Construction}
\begin{justify}

To achieve end-to-end collaborative optimization of the aforementioned calibration, routing, and distillation processes, AARME-GNN incorporates main fusion classification supervision, unimodal auxiliary supervision, routing-aware distillation constraint, and routing entropy regularization into a unified training objective. Specifically, the main classification loss ensures that the fused representation has direct discriminative capability for DDI event prediction, and the unimodal auxiliary loss maintains the independent discriminability of each modality branch. The distillation loss uses routing weights to guide the transfer of modality-specific knowledge to the main fusion classifier, while the routing entropy regularization constrains MMWR to avoid prematurely degenerating into single-modality selection. The main classification loss corresponding to the main fusion classifier is defined using the logits $\textit{o}_{ij}^{M}$ produced by the main fusion classifier. The loss is computed as
\begin{equation}
\mathcal{L}_{\mathrm{main}} = \frac{1}{|\mathcal{B}|} \sum_{(i,j) \in \mathcal{B}}\operatorname{CE}(\textit{o}_{ij}^{M}, y_{ij}).
\end{equation}

By combining the main classification loss, unimodal auxiliary loss, distillation loss, and routing entropy regularization term, the joint optimization objective of the model is formulated as
\begin{equation}
\mathcal{L}_{\mathrm{total}} =
\mathcal{L}_{\mathrm{main}}
+ \lambda_{\mathrm{uni}} \mathcal{L}_{\mathrm{uni}}
+ \lambda_{\mathrm{dist}} \mathcal{L}_\mathrm{dist}
+ \lambda_{\mathrm{balance}} \mathcal{L}_\mathrm{ent},
\end{equation}
where $\lambda_{\mathrm{uni}}$, $\lambda_{\mathrm{dist}}$, and $\lambda_{\mathrm{balance}}$ are penalty coefficients used to balance different training objectives. Specifically, $\lambda_{\mathrm{uni}}$ controls the strength of unimodal auxiliary supervision, which encourages each modality branch to maintain independent discriminative capability. $\lambda_{\mathrm{dist}}$ controls the routing-aware distillation constraint, enabling the main fusion classifier to absorb modality-specific knowledge from the dynamically weighted unimodal predictions. $\lambda_{\mathrm{balance}}$ controls the routing entropy regularization term, which prevents the modality router from collapsing to a single dominant modality and promotes more balanced use of multimodal information. The three coefficients balance prediction accuracy, modality-specific knowledge preservation and routing stability. The model is trained end-to-end with the Adam optimizer. All four terms are averaged over the training batch: the cross-entropy terms use the mean reduction, the distillation term the batch mean of the Kullback--Leibler divergence, and the entropy term the batch mean of the per-sample routing entropies.
\end{justify}

\section*{Results}

\subsection*{Task Description}
\begin{justify}

We evaluate AARME-GNN under three prediction settings of increasing difficulty:

\textbf{Task 1, known-drug DDI prediction.} Both training and test sets contain only drugs that appear in the knowledge graphs, so all drugs are seen during training.

\textbf{Task 2, known--new drug DDI prediction.} The training set contains all known drugs, whereas the test set introduces a subset of previously unseen ``new'' drugs, and the model predicts interactions between known and new drugs.

\textbf{Task 3, new-drug DDI prediction.} The training set again contains only known drugs, and every test pair consists of two new drugs that are completely unseen during training. The task evaluates prediction under cold-start conditions.

\textbf{Representation of unseen drugs.} The unseen drugs of Tasks~2 and 3 have no labeled interactions in the training set, so their representations are constructed entirely from known drugs. For each unseen drug and each of the four modalities we compute its similarity to every training drug, using Jaccard similarity on the modality feature vectors for the chemical-entity, substructure and molecular-structure modalities and an agreement count over the drug--drug relation features for the drug--drug modality. We then select the most similar training drugs, explicitly excluding all drugs of the test fold and the drug itself. The selected training drugs serve as the neighbourhood from which the graph encoder builds the representation of the unseen drug, so that a new drug's representation never originates from test drugs. The five-fold partition of Tasks~2 and 3 is defined at the level of \emph{drugs}, with no drug occurring in both the training and the test set, whereas Task~1 partitions drug \emph{pairs} with stratification over event types.
\end{justify}

\subsection*{Experimental Setup}

\subsubsection*{Dataset}
\begin{justify}

\textbf{Dataset 1, the primary benchmark.} We adopt the four-part drug knowledge graph released with MKG-FENN \cite{wu2024mkg}, which integrates multiple public resources into four knowledge graphs of $<drug, relation, entity>$ triplets. The four parts are constructed as follows.

\textbf{Part 1, drug--chemical entity.} Based on DDI event data, relations between drugs and chemical entities such as transporters and targets were extracted from DrugBank version 5.1.7, and relation labels were assigned according to the general functions of the entities.

\textbf{Part 2, drug--substructure.} Drug SMILES strings provided in DDIMDL \cite{deng2020multimodal} were used as entities, with relations uniformly labeled as ``including'', constructing the knowledge graph between drugs and molecular substructures.

\textbf{Part 3, drug--drug.} Using the DDI event matrix from DDIMDL \cite{deng2020multimodal}, each interacting drug pair was treated as an entity, with specific event types such as ``serum concentration increase'' defined as relations. Following the convention of prior work on this benchmark, DDI events are treated as undirected. A drug pair $(d_i, d_j)$ and its reverse $(d_j, d_i)$ share the same event label.

\textbf{Part 4, molecular structure.} Using the Molecular ACCess System (MACCS) keys \cite{baranwal2020deep}, we selected 13 MACCS keys and 7 additional molecular features and used the frequency of each substructure in a drug as the relation value.

The number of drugs, entities, relations and triplets in each subgraph matches the original release. We use the four released knowledge graphs as the input structure and initialize all drug, relation and entity embeddings randomly, training them end-to-end with the model. The data splits follow the released implementation so that comparisons with methods developed on this benchmark remain like-for-like. The four graphs are not filtered by the train/test split, so the neighbourhood of a training drug may in principle contain edges of test drugs whose embeddings are then updated during training. This transductive property is inherent to the benchmark and is shared by the compared methods. The representation used to predict an unseen drug is nevertheless built only from training drugs, as described above.

\textbf{Dataset 2, the second benchmark.} To test generalization beyond a single inherited benchmark, we additionally use the public large-scale DDI dataset distributed with the MDDI-SCL and MDF-SA-DDI studies, which contains 1{,}258 drugs, 100 DDI event types and 323{,}539 drug pairs. As this benchmark is distributed as flat interaction tables rather than as a ready-made four-part knowledge graph, we constructed its four parts following the design of Dataset~1.

\textbf{Part 1, drug--chemical entity.} Built from the drug target and enzyme annotations of this benchmark, it contains 11{,}597 triples, 1{,}961 entities and two relation types.

\textbf{Part 2, drug--substructure.} Encodes the 2{,}048-bit molecular fingerprint of each drug as substructure entities linked by an ``include'' relation, giving 58{,}436 triples and 2{,}040 entities.

\textbf{Part 3, drug--drug.} Taken directly from the interaction table, it contains 323{,}539 triples over 100 event types.

\textbf{Part 4, molecular structure.} Aggregates the fingerprint bits into 20 binned molecular descriptors and 25{,}160 triples.

The chemical-entity part of Dataset~2 is therefore coarser than that of Dataset~1, which is released as a ready-made knowledge graph with 235 relation types, and the transductive property described above applies to the constructed parts as well. The resulting knowledge graphs are provided in the Supporting Information, and the same five-fold split and evaluation metrics are used as for Dataset~1.

\textbf{Drug pair order and symmetry.} DDI events are undirected, so $(d_i, d_j)$ and $(d_j, d_i)$ carry the same event label, whereas the routing input of MMWR is order-dependent, as described in the \emph{Multimodal Weight Router} subsection. \textit{Training:} every training pair is duplicated with the two drug columns swapped, under the same label. This doubles the training set and exposes the router to both orders. The auxiliary and distillation branches, which are conditioned on the first drug of the ordered pair, see both orders as well. This augmentation is applied in all three prediction tasks and is the one used by MKG-FENN \cite{wu2024mkg}, so that the comparison against that baseline remains like-for-like. \textit{Inference:} each test pair is predicted once, in the order in which it is stored in the dataset, and the predictions of the two directions are not averaged, again following the evaluation convention of MKG-FENN. The code implementing this procedure is provided in S1 File.\end{justify}

\subsubsection*{Evaluation Metrics}
\begin{justify}

Every model is scored with six metrics on the pooled test predictions: accuracy (ACC), precision, recall, the F1 score, AUC and the area under the precision--recall curve (AUPR). These metrics were calculated following the same definitions as those used in MKG-FENN and its Supplementary File \cite{wu2024mkg}. Briefly, ACC measures the overall proportion of correctly predicted samples; precision reflects the proportion of correctly predicted positive samples among all samples predicted as positive; recall measures the proportion of correctly identified positive samples among all true positive samples; and F1 is the harmonic mean of precision and recall. AUC evaluates the area under the receiver operating characteristic curve, constructed from the false-positive and true-positive rates, whereas AUPR evaluates the area under the Precision--Recall curve. Because the two benchmarks contain 65 and 100 event classes, precision, recall and F1 are macro-averaged over the event types, whereas ACC, AUPR and AUC are micro-averaged over all drug pairs. In the tables, precision and recall are abbreviated as Pre and Rec.

For the model comparison we report Win/Tie/Loss (W/T/L) statistics, which count the evaluation metrics on which AARME-GNN achieves a higher, equal or lower value than a given baseline. Two values count as a tie when they agree at the four decimal places reported. For the AARME-GNN row, the counts are summed over all baselines of the setting. The paired test is the two-sided Wilcoxon signed-rank test over the six aggregate metrics ($n=6$, or $n=5$ when one metric is tied), and the Friedman mean rank (F-rank, higher is better) summarizes the ranking of the models across the metrics. In the event-level analysis, each paired observation is the per-event value of one DDI event type and the two-sided paired Wilcoxon signed-rank test is applied metric by metric, with the Holm--Bonferroni correction across the metrics within a task. In Tables~4 and~6, $\Delta$ denotes the difference between AARME-GNN and MKG-FENN. The best value in each column of the comparison tables is boldfaced, and the italic line above each task block of the event-level tables names the metrics that remain significant after the correction.
\end{justify}

\subsubsection*{Baseline Models}
\begin{justify}

The proposed AARME-GNN model is compared with seven representative methods, including MKG-FENN \cite{wu2024mkg}, MDDI-SCL \cite{lin2022mddi}, MDF-SA-DDI \cite{lin2022mdf}, DDIMDL \cite{deng2020multimodal}, MDNN \cite{lyu2021mdnn}, the method by Lee et al. \cite{lee2019novel}, and DeepDDI \cite{ryu2018deep}. Additionally, several conventional classifiers are included as baselines, such as Deep Neural Networks (DNN), Random Forests (RF), k-Nearest Neighbors (KNN), Logistic Regression (LR) \cite{deng2020multimodal}, and Graph Neural Networks (GNN) \cite{kipf2016semi}. A brief description of each model is provided below:

\textbf{MKG-FENN} \cite{wu2024mkg}: A multimodal knowledge graph neural network that constructs four types of KGs---Drug--Chemical Entity, Drug--Substructure, Drug--Drug, and Molecular Structure---and extracts high-order structural and semantic features via a four-channel GNN, followed by end-to-end fusion through an MLP for DDI prediction.

\textbf{MDDI-SCL} \cite{lin2022mddi}: A supervised contrastive learning-based method with a three-level loss framework for multi-type DDI prediction. It combines drug feature encoding with mean-squared error, latent feature fusion with contrastive loss for drug pairs, and multi-type DDI prediction with classification loss.

\textbf{MDF-SA-DDI} \cite{lin2022mdf}: Predicts DDI events by integrating multi-source drug features and employing a Transformer-based self-attention mechanism. It combines latent features from four distinct drug fusion networks using a Transformer.

\textbf{DDIMDL} \cite{deng2020multimodal}: A multimodal deep learning framework that leverages four drug features—chemical substructures, targets, enzymes, and pathways—by constructing separate deep neural submodels and jointly learning cross-modal representations for DDI event prediction.

\textbf{MDNN} \cite{lyu2021mdnn}: A dual-path multimodal deep neural network combining a drug knowledge graph pathway and a heterogeneous feature pathway. Topological and semantic relations are extracted using GNNs and fused to predict DDI events.

\textbf{Lee et al.'s method} \cite{lee2019novel}: Uses autoencoder-based dimensionality reduction and deep feedforward networks to predict DDI effect types based on chemical structure similarity spectra, gene ontology similarity spectra, and target gene similarity spectra.

\textbf{DeepDDI} \cite{ryu2018deep}: A deep neural network-based method that predicts 86 types of drug interactions using principal component analysis on drug chemical structure similarity matrices.

\textbf{Conventional classifiers} \cite{deng2020multimodal,kipf2016semi}: Includes DNN, RF, KNN, LR, and GNN, which are evaluated as additional baseline methods.

Four recent methods are not directly comparable with our multi-class results. DGNN-DDI~\cite{ma2023dual} scores each interaction type separately, with the type as an input of a binary function $f(d_x,d_y,r)\rightarrow\{0,1\}$, so its output is a per-type decision rather than a multi-class prediction. TIGER \cite{su2024dual} and KA-DDI \cite{li2026ka} predict whether an interaction exists between two drugs, and MeTDDI \cite{zhong2024learning} predicts directional perpetrator--victim roles and changes in drug plasma concentration. Because none of these formulations produces a calibrated multi-class decision over event types, the four methods are not included in the comparison tables.

All baseline models use the hyperparameters recommended in their original publications. The metrics of the compared methods are those reported in the original studies, and the model that we re-ran ourselves, MKG-FENN on Dataset~2, was evaluated under our own five-fold split. The main hyperparameters of the AARME-GNN model are set as follows: the number of training epochs is 120, the batch size is 1024, weight decay is set to $1\times10^{-8}$, and the learning rate is $1\times10^{-2}$. The neighbor sampling size is 6, the number of DDI event classes is 65, the total number of drugs is 572, and the embedding dimensionality is 128. The configuration of Tasks~2 and 3 follows the released implementation of MKG-FENN~\cite{wu2024mkg} for those two settings, so that the two models are compared under identical conditions. These tasks use three of the four modality branches, the molecular-structure part being used in Task~1 only, an embedding dimensionality of 256 in place of 128, a learning rate of $5\times10^{-3}$ in place of $1\times10^{-2}$, and 100 training epochs on Task~2 in place of 120, whereas Task~3 keeps 120. All remaining settings are shared by the three tasks: batch size 1024, weight decay $1\times10^{-8}$, neighbor sampling size 6, dropout 0.3, bottleneck dimension 64, and loss coefficients $\lambda_{\mathrm{balance}}=0.2$, $\lambda_{\mathrm{dist}}=0.1$, $\lambda_{\mathrm{uni}}=0.1$. A sensitivity analysis of the auxiliary-head input, raw versus FACM-calibrated representation, is provided in S3 Table. Dataset~1 experiments were run on a workstation with an Intel Core Ultra 7 155H processor at 3.8 GHz and 32 GB of RAM under Windows~11, and the Dataset~2 experiments on a GPU-equipped server.
\end{justify}

\subsection*{Comparison on Dataset 1}

\subsubsection*{Comparison on Known Drugs (Task 1)}
\begin{figure}[!ht]
\begin{adjustwidth}{-4cm}{0cm}
\centering
\includegraphics[height=3.918in]{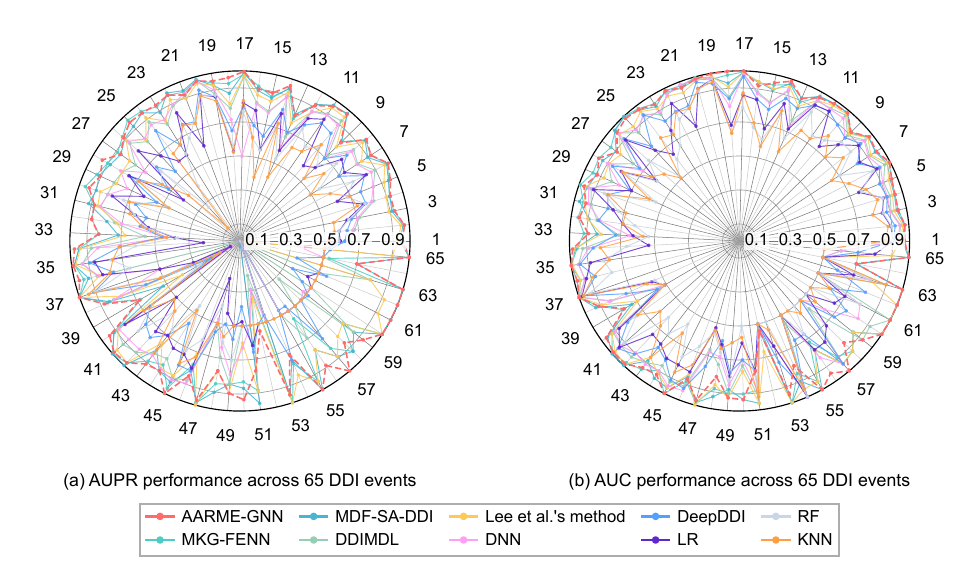}
\captionsetup{width=1.3\textwidth}
\caption{\bf Event-wise AUPR and AUC scores of different prediction models.}
\label{fig:task1_eventwise_aupr_auc}
\end{adjustwidth}
\end{figure}

\begin{figure}[!ht]
\begin{adjustwidth}{-5cm}{0cm}
\centering
\includegraphics[height=2.527in]{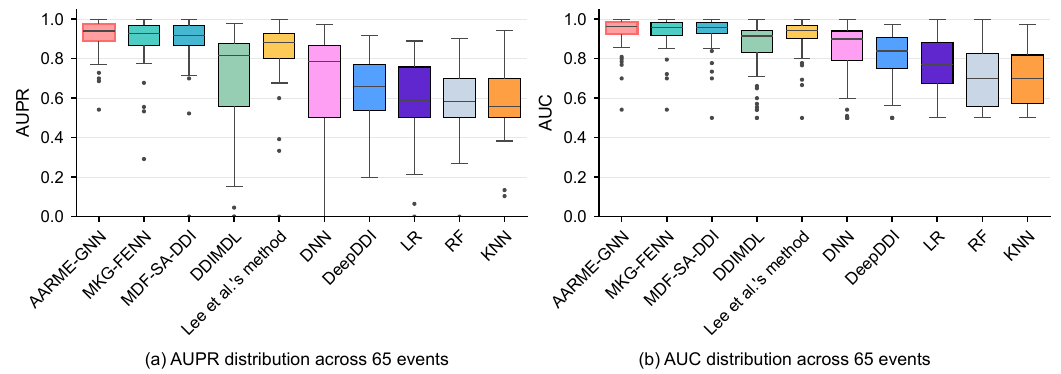}
\captionsetup{width=1.3\textwidth}
\caption{\bf Boxplots of AUPR and AUC across DDI events for different methods.}
\label{fig:task1_aupr_auc_boxplots}
\end{adjustwidth}
\end{figure}

\begin{justify}

Task 1 is DDI event prediction between known drugs, the standard closed-world setting. As Table~\ref{tab:task1_shifted} shows, AARME-GNN attains the highest accuracy, AUC, F1 and precision. MKG-FENN is slightly higher on AUPR and recall. AARME-GNN reaches an accuracy of 0.9431 against 0.9409 for MKG-FENN, a gain of 0.22 percentage points, a macro-averaged F1 of 0.9048 against 0.8958, and a macro-averaged precision of 0.9503 against 0.9132, an improvement of 3.71 percentage points. The margin over MKG-FENN on this task is therefore modest, as expected because both models operate close to saturation. The precision gain is accompanied by a slightly lower recall, 0.8824 against 0.8876, and the paired test over the six aggregate metrics does not reach significance. Across the twelve baselines AARME-GNN wins 69 metric comparisons and loses 3.

AARME-GNN attains the highest or second-highest AUPR and AUC for most event classes, as Fig~\ref{fig:task1_eventwise_aupr_auc} shows. For the events with the fewest pairs, such as events 45 and 56, AARME-GNN keeps high accuracy where the baselines rarely predict the event at all. The boxplots in Fig~\ref{fig:task1_aupr_auc_boxplots} show tighter AUPR and AUC distributions for AARME-GNN, with smaller interquartile ranges across the 65 events. The statistical comparison of the models is reported in the \emph{Event-Level and Metric-Level Significance Analysis} subsection.

\end{justify}

\begin{table}[!ht]
\begin{adjustwidth}{-4cm}{0cm}
\captionsetup{justification=raggedright, singlelinecheck=false}
\normalsize
\setlength{\tabcolsep}{4pt}
\renewcommand{\arraystretch}{1.2}
\caption{\bf Performance comparison between AARME-GNN and the competing models in Task~1 on Dataset~1.}
\label{tab:task1_shifted}
\begin{tabular*}{\linewidth}{@{\extracolsep{\fill}}l c c c c c c c c c@{}}
\toprule
\multicolumn{10}{c}{\textbf{Task 1}} \\
\midrule
\textbf{Model} & \textbf{ACC} & \textbf{AUPR} & \textbf{AUC} & \textbf{F1} & \textbf{Pre} & \textbf{Rec} & \textbf{Win/Tie/Loss} & \textbf{$p$} & \textbf{F-rank} \\
\midrule
MKG-FENN             & 0.9409 & \textbf{0.9786} & 0.9989 & 0.8958 & 0.9132 & \textbf{0.8876} & 4/0/2 & 0.5625 & 12.17 \\
MDDI-SCL             & 0.9378 & 0.9782 & 0.9983 & 0.8755 & 0.8804 & 0.8767 & 5/0/1 & 0.09375 & 10.17 \\
MDF-SA-DDI           & 0.9301 & 0.9737 & 0.9989 & 0.8878 & 0.9085 & 0.8760 & 6/0/0 & 0.03125 & 10.50 \\
DDIMDL              & 0.8852 & 0.9208 & 0.9976 & 0.7585 & 0.8471 & 0.7182 & 6/0/0 & 0.03125 & 6.17 \\
MDNN                 & 0.9175 & 0.9668 & 0.9984 & 0.8301 & 0.8622 & 0.8202 & 6/0/0 & 0.03125 & 8.17 \\
Lee et al.  & 0.9094 & 0.9562 & 0.9961 & 0.8391 & 0.8509 & 0.8339 & 6/0/0 & 0.03125 & 7.25 \\
DeepDDI              & 0.8371 & 0.8899 & 0.9961 & 0.6848 & 0.7275 & 0.6611 & 6/0/0 & 0.03125 & 3.75 \\
DNN                  & 0.8797 & 0.9134 & 0.9963 & 0.7223 & 0.8047 & 0.7027 & 6/0/0 & 0.03125 & 5.17 \\
RF                   & 0.7775 & 0.8349 & 0.9956 & 0.5936 & 0.7893 & 0.5161 & 6/0/0 & 0.03125 & 2.33 \\
KNN                  & 0.7214 & 0.7716 & 0.9813 & 0.4831 & 0.7174 & 0.4081 & 6/0/0 & 0.03125 & 1.00 \\
LR                   & 0.7920 & 0.8400 & 0.9960 & 0.5948 & 0.7437 & 0.5236 & 6/0/0 & 0.03125 & 3.00 \\
GNN                  & 0.9142 & 0.9691 & 0.9989 & 0.8332 & 0.8941 & 0.8012 & 6/0/0 & 0.03125 & 8.83 \\
AARME-GNN            & \textbf{0.9431} & 0.9751 & \textbf{0.9992} & \textbf{0.9048} & \textbf{0.9503} & 0.8824 & \textbf{69/0/3} & --- & \textbf{12.50} \\
\bottomrule
\end{tabular*}
\end{adjustwidth}
\end{table}

\subsubsection*{Comparison on New Drugs (Task 2 and 3)}
\begin{justify}

Tasks 2 and 3 test the model on new drugs. The first setting uses pairs of one known and one new drug, and the second uses pairs of two new drugs. Tables~\ref{tab:task2_comparison} and~\ref{tab:task3_comparison} give the comparisons between AARME-GNN and the baseline models on Task 2 and 3, respectively.

In Task 2, AARME-GNN achieves the highest AUC, 0.9828, and the highest macro-averaged precision, 0.6720, of all compared models. Its AUC exceeded the 0.9673 of MKG-FENN, and its precision exceeded the 0.6063 of MKG-FENN by 6.57 percentage points, indicating more reliable positive predictions. MKG-FENN remains marginally ahead in accuracy (0.6805 against 0.6789) and AUPR (0.7049 against 0.7011), and the paired test over the six metrics does not reach significance, with $p=0.6875$. At the event level, however, AARME-GNN is significantly better than MKG-FENN in accuracy, F1 and precision, with Holm-corrected $p\le0.017$. The Win/Tie/Loss statistics over the 42 metric-level comparisons against the seven baselines give 36 wins and 6 losses. This pattern reflects the same precision--recall trade-off observed in Task 1, with a recall 2.46 percentage points below that of MKG-FENN, 0.4860 against 0.5106: the model makes fewer but more reliable positive predictions.

On the harder Task 3, AARME-GNN also shows advantages on several metrics. Its accuracy reaches 0.4743 against 0.4552 for MKG-FENN, its AUPR 0.4465 against 0.4162, and its AUC 0.9602 against 0.9149. Although the absolute scores are low because predicting interactions between completely unseen drugs is difficult, AARME-GNN achieves the highest accuracy, AUPR and AUC of all compared models, as well as the highest precision, 0.3062 against 0.2754 for MKG-FENN. This is a gain of 3.08 percentage points in precision at a recall 2.00 percentage points lower, 0.1931 against 0.2131. This pattern indicates that the proposed mechanisms contribute mainly to the precision of predictions under distribution shift, while recall in the cold-start settings remains an area for improvement.
\end{justify}

\begin{table}[!ht]
\begin{adjustwidth}{-4.3cm}{0cm}
\captionsetup{justification=raggedright, singlelinecheck=false}
\normalsize
\setlength{\tabcolsep}{4pt}
\renewcommand{\arraystretch}{1.2}
\caption{\bf Performance comparison between AARME-GNN and the competing models in Task~2 on Dataset~1.}
\label{tab:task2_comparison}
\begin{tabular*}{\linewidth}{@{\extracolsep{\fill}}l c c c c c c c c c@{}}
\toprule
\multicolumn{10}{c}{\textbf{Task 2}} \\
\midrule
\textbf{Model} & \textbf{ACC} & \textbf{AUPR} & \textbf{AUC} & \textbf{F1} & \textbf{Pre} & \textbf{Rec} & \textbf{Win/Tie/Loss} & \textbf{$p$} & \textbf{F-rank} \\
\midrule
\textbf{MKG-FENN}      & \textbf{0.6805} & \textbf{0.7049} & 0.9673 & 0.5394 & 0.6063 & \textbf{0.5106} & 3/0/3 & 0.6875 & 6.67 \\
\textbf{MDDI-SCL}      & 0.6767 & 0.6947 & 0.9634 & 0.5304 & 0.6254 & 0.4814 & 6/0/0 & 0.03125 & 5.17 \\
\textbf{MDF-SA-DDI}    & 0.6633 & 0.6776 & 0.9497 & \textbf{0.5584} & 0.6547 & 0.5078 & 4/0/2 & 0.3125 & 5.67 \\
\textbf{DDIMDL}       & 0.6415 & 0.6558 & 0.9799 & 0.4460 & 0.5607 & 0.4319 & 6/0/0 & 0.03125 & 4.17 \\
\textbf{Lee et al.} & 0.6405 & 0.6244 & 0.9247 & 0.5039 & 0.5388 & 0.4891 & 5/0/1 & 0.0625 & 3.17 \\
\textbf{DeepDDI}       & 0.5774 & 0.5594 & 0.9575 & 0.3416 & 0.3630 & 0.3890 & 6/0/0 & 0.03125 & 1.67 \\
\textbf{DNN}           & 0.6239 & 0.6361 & 0.9796 & 0.2997 & 0.4237 & 0.2840 & 6/0/0 & 0.03125 & 2.50 \\
\textbf{AARME-GNN}     & 0.6789 & 0.7011 & \textbf{0.9828} & 0.5477 & \textbf{0.6720} & 0.4860 & \textbf{36/0/6} & --- & \textbf{7.00} \\
\bottomrule
\end{tabular*}
\end{adjustwidth}
\end{table}

\begin{table}[!ht]
\begin{adjustwidth}{-4.3cm}{0cm}
\normalsize
\setlength{\tabcolsep}{4pt}
\renewcommand{\arraystretch}{1.2}
\captionsetup{justification=raggedright, singlelinecheck=false}
\caption{\bf Performance comparison between AARME-GNN and the competing models in Task~3 on Dataset~1.}
\label{tab:task3_comparison}
\begin{tabular*}{\linewidth}{@{\extracolsep{\fill}}l c c c c c c c c c@{}}
\toprule
\multicolumn{10}{c}{\textbf{Task 3}} \\
\midrule
\textbf{Model} & \textbf{ACC} & \textbf{AUPR} & \textbf{AUC} & \textbf{F1} & \textbf{Pre} & \textbf{Rec} & \textbf{Win/Tie/Loss} & \textbf{$p$} & \textbf{F-rank} \\
\midrule
\textbf{MKG-FENN}      & 0.4552 & 0.4162 & 0.9149 & 0.2186 & 0.2754 & 0.2131 & 4/0/2 & 0.2188 & 6.50 \\
\textbf{MDDI-SCL}      & 0.4589 & 0.3938 & 0.9053 & 0.1919 & 0.2585 & 0.1678 & 6/0/0 & 0.03125 & 4.83 \\
\textbf{MDF-SA-DDI}    & 0.4338 & 0.3873 & 0.8630 & \textbf{0.2329} & 0.2715 & \textbf{0.2226} & 4/0/2 & 0.1562 & 5.67 \\
\textbf{DDIMDL}       & 0.4075 & 0.3635 & 0.9512 & 0.1590 & 0.2408 & 0.1452 & 6/0/0 & 0.03125 & 3.50 \\
\textbf{Lee et al.} & 0.4097 & 0.3184 & 0.8302 & 0.2022 & 0.2216 & 0.2027 & 5/0/1 & 0.09375 & 3.50 \\
\textbf{DeepDDI}       & 0.3602 & 0.2781 & 0.9059 & 0.1373 & 0.1586 & 0.1450 & 6/0/0 & 0.03125 & 1.83 \\
\textbf{DNN}           & 0.4087 & 0.3776 & 0.9550 & 0.1152 & 0.1836 & 0.1093 & 6/0/0 & 0.03125 & 3.00 \\
\textbf{AARME-GNN}     & \textbf{0.4743} & \textbf{0.4465} & \textbf{0.9602} & 0.2100 & \textbf{0.3062} & 0.1931 & \textbf{37/0/5} & --- & \textbf{7.17} \\
\bottomrule
\end{tabular*}
\end{adjustwidth}
\end{table}

\subsubsection*{Event-Level and Metric-Level Significance Analysis}
\begin{justify}

The paired observations of the event-level tests are the per-event values of the
65 DDI event types of Dataset~1, which are the values plotted for Task~1 in Fig~\ref{fig:task1_eventwise_aupr_auc}. Each model contributes one observation per event type. The aggregate AUPR in Table~\ref{tab:task1_shifted} is the micro
average, weighted by the number of drug pairs and therefore dominated by the most
frequent event types, whereas the signed-rank test treats all event types
equally, so the two quantities can point in different directions. On this
benchmark the aggregate AUPR of MKG-FENN is slightly higher than that of
AARME-GNN, 0.9786 against 0.9751, and the difference over the six aggregate
metrics is not significant ($p=0.5625$). No AUPR advantage is therefore claimed.

The event-level comparisons are given in Table~\ref{tab:eventlevel_sig_d1}. On
Task~1, AARME-GNN is significantly better on accuracy
($p_{\mathrm{Holm}}=0.0050$) and precision ($p_{\mathrm{Holm}}=0.0110$), whereas
F1 and recall do not differ significantly. On Task~2, accuracy, F1 and precision
are significantly higher and recall is significantly lower. On Task~3, accuracy is significantly higher and recall is significantly lower. AUPR and AUC are not
included in these tests, because their per-event values are shown in
Fig~\ref{fig:task1_eventwise_aupr_auc} and their significance is assessed at
the metric level in Table~\ref{tab:task1_shifted}.

Under the metric-level tests, AARME-GNN attains the highest F-rank on all three tasks, 12.50, 7.00 and 7.17. The Friedman test rejects the null hypothesis of equivalent performance on every task, with $\chi^2=63.505$ ($p=2.29\times10^{-12}$) for Task~1, $\chi^2=37.643$ ($p=4.45\times10^{-7}$) for Task~2 and $\chi^2=39.143$ ($p=2.22\times10^{-7}$) for Task~3.

\end{justify}

\begin{table}[!ht]
\centering
\small
\setlength{\tabcolsep}{6pt}
\renewcommand{\arraystretch}{1.05}
\captionsetup{justification=raggedright, singlelinecheck=false}
\caption{\bf Event-level paired comparisons against MKG-FENN on Dataset~1.}
\label{tab:eventlevel_sig_d1}
\begin{tabular*}{\linewidth}{@{\extracolsep{\fill}}l r r r r@{}}
\toprule
\textbf{Metric} & \textbf{AARME-GNN} & \textbf{MKG-FENN} & \textbf{$\Delta$} & \textbf{$p_{\mathrm{Holm}}$} \\
\midrule
\multicolumn{5}{@{}c}{\textbf{Task~1}}\\
\cmidrule(l){1-5}
ACC & \textbf{0.9982} & 0.9981 & $+0.0001$ & 0.00502 \\
F1 & \textbf{0.9048} & 0.9005 & $+0.0043$ & 0.349 \\
Pre & \textbf{0.9503} & 0.9350 & $+0.0153$ & 0.0110 \\
Rec & 0.8824 & \textbf{0.8858} & $-0.0034$ & 0.489 \\
\cmidrule(l){1-5}
\multicolumn{5}{@{}c}{\textbf{Task~2}}\\
\cmidrule(l){1-5}
ACC & \textbf{0.9901} & 0.9900 & $+0.0001$ & $4.55\times10^{-4}$ \\
F1 & \textbf{0.5477} & 0.5324 & $+0.0152$ & 0.0168 \\
Pre & \textbf{0.6720} & 0.5889 & $+0.0831$ & $1.60\times10^{-6}$ \\
Rec & 0.4860 & \textbf{0.5073} & $-0.0212$ & 0.00214 \\
\cmidrule(l){1-5}
\multicolumn{5}{@{}c}{\textbf{Task~3}}\\
\cmidrule(l){1-5}
ACC & \textbf{0.9838} & 0.9828 & $+0.0011$ & 0.0153 \\
F1 & 0.2100 & \textbf{0.2211} & $-0.0112$ & 0.0665 \\
Pre & \textbf{0.3062} & 0.2837 & $+0.0225$ & 0.249 \\
Rec & 0.1931 & \textbf{0.2171} & $-0.0240$ & 0.0278 \\
\bottomrule
\end{tabular*}
\end{table}

\subsection*{Comparison on Dataset 2}
\begin{justify}

To test whether the behavior of AARME-GNN extends beyond the benchmark on which
it was developed, we repeated the three prediction settings on Dataset~2 with the same five-fold split and the same six evaluation metrics. We ran AARME-GNN and MKG-FENN on the identical split. The remaining methods are quoted from the results published by MDDI-SCL~\cite{lin2022mddi} for this benchmark and were not re-run, so those rows serve as reference values rather than as a controlled head-to-head comparison. RF, KNN and LR were evaluated in Task~1 only in that study. Table~\ref{tab:dataset2_comparison} reports
the results for the three settings.

On Task~1 the two models we ran are close. MKG-FENN is marginally ahead in accuracy, AUPR, F1 and recall, scoring 0.9513 against 0.9479, 0.9877 against 0.9835, 0.9402 against 0.9360 and 0.9411 against 0.9295. The event-level tests favor MKG-FENN on the same four metrics and on AUC, whereas the difference in precision is not significant. AARME-GNN attains the highest precision of all compared methods, 0.9455 against 0.9405 for MKG-FENN and 0.9381 for MDF-SA-DDI. The
advantage of AARME-GNN over MKG-FENN appears in the two cold-start settings. On Task~2 it is better on five of the six metrics, with accuracy 0.6732 against 0.6457, AUPR 0.6920 against 0.6707, AUC 0.9873 against 0.9821, F1 0.5507 against 0.5221 and precision 0.6616 against 0.5747. The event-level paired tests show that the gains in accuracy, AUPR, F1 and precision are significant, with Holm-corrected $p<10^{-10}$ (Table~\ref{tab:eventlevel_sig}). In these tests AUPR and AUC are computed from
the predicted probability matrices, and in Task~2 they use 99 event types
because one type has no positive sample in one fold. On Task~3 it attains the highest accuracy and
AUPR of all compared methods, 0.4925 and 0.4603, and is significantly better than
MKG-FENN on accuracy, AUPR and precision, with Holm-corrected $p<10^{-4}$. The precision--recall trade-off observed on Dataset~1 reappears here. MKG-FENN keeps a small recall advantage on Task~1, and the published baselines obtain the best individual scores on some metrics of Task~2 and~3, such as Lee et al., who reach the highest accuracy and AUPR on Task~2. No single method dominates this benchmark.

At the metric level the comparison follows the same procedure as for Dataset~1: the
paired unit is the six metrics, and we report Win/Tie/Loss counts, the paired
Wilcoxon signed-rank test and the Friedman F-rank. Against the direct
predecessor, AARME-GNN attains the higher F-rank in both cold-start settings,
5.17 against 2.83 on Task~2 and 5.67 against 2.50 on Task~3, and wins five of the
six metrics in each, whereas MKG-FENN leads on Task~1 with an F-rank of 10.42
against 9.75. No single method attains the highest F-rank in every setting:
MKG-FENN leads on Task~1 and DDIMDL on Task~2 and~3, and the Friedman test
rejects the null hypothesis of equivalent performance in all three settings.
AARME-GNN wins 52 of its 60 metric-level comparisons on Task~1 and 25 and 28 of
42 on Task~2 and~3, with the losses in the cold-start settings concentrated in
recall and F1.

The absolute scores on these two settings are low for every compared method, with
accuracies between 0.36 and 0.69. This reflects the difficulty of the setting
rather than a property of a particular model. The new drugs of Tasks~2 and~3 have no labeled interactions in the training set, so their representations are inferred from the most similar known drugs. The event types of this benchmark are unevenly represented, with the 100 types ranging from 51 to 100{,}725 drug pairs and the ten most frequent accounting for 82\% of all pairs. The macro-averaged
scores therefore weight rare event types as heavily as frequent ones, while the
event-level tests still detect significant differences within this range.

The second benchmark reproduces the pattern of Dataset~1:
AARME-GNN is competitive in the closed-world setting, and its advantage over the
direct predecessor is concentrated in the cold-start settings and in precision
rather than being uniform across metrics. The consistency of this pattern across
two benchmarks that differ in scale, event vocabulary and knowledge-graph
construction supports the generalization of the proposed components.

\end{justify}

\begin{table}[!htbp]
\begin{adjustwidth}{-4.3cm}{0cm}
\normalsize
\setlength{\tabcolsep}{4pt}
\renewcommand{\arraystretch}{1.2}
\captionsetup{justification=raggedright, singlelinecheck=false}
\caption{\bf Performance comparison on Dataset~2 under the three prediction settings.}
\label{tab:dataset2_comparison}
\renewcommand{\arraystretch}{1.1}
\begin{tabular*}{\linewidth}{@{\extracolsep{\fill}}l c c c c c c c c c@{}}
\toprule
\textbf{Model} & \textbf{ACC} & \textbf{AUPR} & \textbf{AUC} & \textbf{F1} & \textbf{Pre} & \textbf{Rec} & \textbf{Win/Tie/Loss} & \textbf{$p$} & \textbf{F-rank} \\
\midrule
\multicolumn{10}{c}{\textbf{Task 1}} \\
\midrule
MKG-FENN & 0.9513 & \textbf{0.9877} & \textbf{0.9997} & \textbf{0.9402} & 0.9405 & 0.9411 & 1/1/4 & 0.4375 & \textbf{10.42} \\
MDDI-SCL & \textbf{0.9516} & 0.9862 & 0.9995 & 0.9321 & 0.9162 & \textbf{0.9500} & 3/0/3 & 1 & 9.17 \\
MDF-SA-DDI & 0.9291 & 0.9773 & 0.9996 & 0.9117 & 0.9381 & 0.8910 & 6/0/0 & 0.03125 & 7.50 \\
DDIMDL & 0.9229 & 0.9637 & 0.9993 & 0.9105 & 0.9212 & 0.9039 & 6/0/0 & 0.03125 & 6.50 \\
Lee et al. & 0.9370 & 0.9791 & 0.9991 & 0.9181 & 0.9226 & 0.9153 & 6/0/0 & 0.03125 & 7.67 \\
DeepDDI & 0.7211 & 0.7724 & 0.9914 & 0.6854 & 0.6654 & 0.7183 & 6/0/0 & 0.03125 & 3.83 \\
DNN & 0.7908 & 0.8539 & 0.9949 & 0.7649 & 0.7560 & 0.8046 & 6/0/0 & 0.03125 & 5.00 \\
RF & 0.6956 & 0.7567 & 0.9892 & 0.5760 & 0.6694 & 0.5426 & 6/0/0 & 0.03125 & 3.17 \\
KNN & 0.5797 & 0.5964 & 0.8998 & 0.3805 & 0.4758 & 0.3347 & 6/0/0 & 0.03125 & 1.83 \\
LR & 0.5229 & 0.5288 & 0.9805 & 0.2373 & 0.3128 & 0.2185 & 6/0/0 & 0.03125 & 1.17 \\
AARME-GNN & 0.9479 & 0.9835 & \textbf{0.9997} & 0.9360 & \textbf{0.9455} & 0.9295 & \textbf{52/1/7} & --- & 9.75 \\
\midrule
\multicolumn{10}{c}{\textbf{Task 2}} \\
\midrule
MKG-FENN & 0.6457 & 0.6707 & 0.9821 & 0.5221 & 0.5747 & 0.4916 & 5/0/1 & 0.0625 & 2.83 \\
MDDI-SCL & 0.6595 & 0.6794 & 0.9757 & 0.5578 & 0.5605 & 0.5712 & 4/0/2 & 0.4375 & 3.67 \\
MDF-SA-DDI & 0.6664 & 0.6820 & 0.9862 & 0.5919 & 0.6526 & 0.5518 & 4/0/2 & 1 & 5.17 \\
DDIMDL & 0.6720 & 0.7086 & \textbf{0.9885} & 0.5817 & 0.6680 & 0.5295 & 1/0/5 & 0.0625 & \textbf{6.17} \\
Lee et al. & \textbf{0.6917} & \textbf{0.7119} & 0.9687 & 0.5934 & 0.6144 & \textbf{0.5848} & 2/0/4 & 0.5625 & 6.00 \\
DeepDDI & 0.5883 & 0.5851 & 0.9746 & 0.4709 & 0.5250 & 0.4361 & 6/0/0 & 0.03125 & 1.17 \\
DNN & 0.6687 & 0.6838 & 0.9818 & \textbf{0.6164} & \textbf{0.7279} & 0.5479 & 3/0/3 & 0.4375 & 5.83 \\
AARME-GNN & 0.6732 & 0.6920 & 0.9873 & 0.5507 & 0.6616 & 0.4885 & \textbf{25/0/17} & --- & 5.17 \\
\midrule
\multicolumn{10}{c}{\textbf{Task 3}} \\
\midrule
MKG-FENN & 0.4509 & 0.4106 & 0.9536 & 0.2358 & 0.3116 & 0.2187 & 5/0/1 & 0.09375 & 2.50 \\
MDDI-SCL & 0.4696 & 0.4261 & 0.9315 & 0.2838 & 0.3160 & 0.2773 & 4/0/2 & 1 & 4.17 \\
MDF-SA-DDI & 0.4794 & 0.4450 & \textbf{0.9686} & 0.2937 & 0.3667 & 0.2659 & 3/0/3 & 0.84375 & 6.00 \\
DDIMDL & 0.4699 & 0.4386 & 0.9685 & 0.3032 & 0.3773 & 0.2729 & 3/0/3 & 0.84375 & \textbf{6.17} \\
Lee et al. & 0.4867 & 0.4349 & 0.9093 & \textbf{0.3082} & 0.3355 & \textbf{0.3066} & 4/0/2 & 1 & 5.50 \\
DeepDDI & 0.3611 & 0.2820 & 0.9264 & 0.1868 & 0.2301 & 0.1711 & 6/0/0 & 0.03125 & 1.17 \\
DNN & 0.4570 & 0.4129 & 0.9565 & 0.2997 & \textbf{0.4345} & 0.2508 & 3/0/3 & 0.5625 & 4.83 \\
AARME-GNN & \textbf{0.4925} & \textbf{0.4603} & 0.9684 & 0.2432 & 0.3825 & 0.2079 & \textbf{28/0/14} & --- & 5.67 \\
\bottomrule
\end{tabular*}
\end{adjustwidth}
\end{table}

\begin{table}[!ht]
\centering
\small
\setlength{\tabcolsep}{6pt}
\renewcommand{\arraystretch}{1.05}
\captionsetup{justification=raggedright, singlelinecheck=false}
\caption{\bf Event-level paired comparisons against MKG-FENN on Dataset~2.}
\label{tab:eventlevel_sig}
\begin{tabular*}{\linewidth}{@{\extracolsep{\fill}}l r r r r@{}}
\toprule
\textbf{Metric} & \textbf{AARME-GNN} & \textbf{MKG-FENN} & \textbf{$\Delta$} & \textbf{$p_{\mathrm{Holm}}$} \\
\midrule
\multicolumn{5}{@{}c}{\textbf{Task~1}}\\
\cmidrule(l){1-5}
ACC & 0.9990 & 0.9990 & $-0.0001$ & $1.94\times10^{-4}$ \\
AUPR & 0.9655 & \textbf{0.9689} & $-0.0034$ & $1.26\times10^{-3}$ \\
AUC & 0.9983 & \textbf{0.9987} & $-0.0004$ & $1.26\times10^{-3}$ \\
F1 & 0.9360 & \textbf{0.9402} & $-0.0042$ & 0.0159 \\
Pre & \textbf{0.9455} & 0.9405 & $+0.0051$ & 0.150 \\
Rec & 0.9295 & \textbf{0.9411} & $-0.0117$ & $1.26\times10^{-3}$ \\
\cmidrule(l){1-5}
\multicolumn{5}{@{}c}{\textbf{Task~2}}\\
\cmidrule(l){1-5}
ACC & \textbf{0.9935} & 0.9929 & $+0.0006$ & $4.75\times10^{-15}$ \\
AUPR & \textbf{0.5278} & 0.4938 & $+0.0339$ & $6.35\times10^{-11}$ \\
AUC & 0.9397 & \textbf{0.9507} & $-0.0110$ & 0.0373 \\
F1 & \textbf{0.5507} & 0.5221 & $+0.0286$ & $5.63\times10^{-11}$ \\
Pre & \textbf{0.6616} & 0.5747 & $+0.0869$ & $1.26\times10^{-16}$ \\
Rec & 0.4885 & \textbf{0.4916} & $-0.0031$ & 0.730 \\
\cmidrule(l){1-5}
\multicolumn{5}{@{}c}{\textbf{Task~3}}\\
\cmidrule(l){1-5}
ACC & \textbf{0.9899} & 0.9890 & $+0.0008$ & $5.00\times10^{-13}$ \\
AUPR & \textbf{0.2381} & 0.2158 & $+0.0223$ & $9.82\times10^{-5}$ \\
AUC & 0.8414 & \textbf{0.8583} & $-0.0168$ & 0.0700 \\
F1 & \textbf{0.2432} & 0.2358 & $+0.0073$ & 0.0700 \\
Pre & \textbf{0.3825} & 0.3116 & $+0.0709$ & $1.45\times10^{-5}$ \\
Rec & 0.2079 & \textbf{0.2187} & $-0.0108$ & 0.163 \\
\bottomrule
\end{tabular*}
\end{table}

\subsection*{Ablation Study}
\begin{justify}

To test what each component contributes, we compare the full model with four variants that remove FACM, RAWD or the routing-based components (Table~\ref{tab:ablation_losses}). The results are reported in Table~\ref{tab:ablation}.

The full model performs best on most metrics: accuracy 0.9431, AUC 0.9992, F1 0.9048, precision 0.9503 and an average of 0.9425 over the six metrics, denoted Avg in Table~\ref{tab:ablation}. The margins over the strongest variant are nevertheless small, at 0.02 percentage points for accuracy, 0.01 for AUC, 0.74 for F1, 1.51 for precision and 0.52 for the average score. For AUPR and recall the full model is not the best configuration, differing by at most 0.32 and 0.01 percentage points. We tested these differences with the same tests used for the main tables. The full model attains the best Friedman mean rank, 4.33 against 2.08--3.17 for the variants, and wins four to six of the six metrics against every variant. The paired signed-rank test reaches significance only for the removal of FACM, with 6/0/0 and $p=0.03125$, whereas the Friedman omnibus test over the five configurations is not significant, with $\chi^2=7.832$ and $p=0.098$. Given the limited power of six paired observations, we interpret the ablation as a consistent trend that supports the contribution of each component rather than as proof that each module is individually essential.

The one metric on which the full model trails by more than a rounding margin is AUPR, where the variant without MMWR and RAWD reaches 0.9783 against 0.9751. The AUPR values of the five configurations span only 0.32 percentage points, and the full model attains the highest precision at 0.9503, so we read this as a shift in the precision--recall operating point rather than as evidence against the full model. Removing FACM leads to a consistent drop in overall performance, with an average score of 0.9373 against 0.9425 and a precision of 0.9352 against 0.9503, indicating that reducing within-modality noise and re-weighting channel responses improves the precision of positive predictions. The variant without RAWD also shows reduced precision and F1 score, demonstrating that routing-aware distillation helps preserve discriminative unimodal knowledge during fusion. In addition, the model without MMWR-based routing performs worse on most metrics, highlighting the benefit of sample-specific modality weighting for adaptive fusion.

\pagebreak[4]
Configurations removing both MMWR and RAWD are not further decomposed, since RAWD is built upon the routing weights produced by MMWR and cannot operate independently. When MMWR is removed, the four modality representations of the two drugs are concatenated and passed through a static multilayer-perceptron fusion layer, so fusion is still learned but no longer sample-adaptive, and the routing entropy term is dropped together with the router. All variants optimize the same joint objective, with the coefficients of the removed components set to zero. The unimodal auxiliary loss $\mathcal{L}_{\mathrm{uni}}$ is retained when RAWD is removed, with only $\lambda_{\mathrm{dist}}$ set to zero. The variant without the router uses the main classification loss only, because without the routing weights the entropy term, the distillation target and the auxiliary branches that support them are removed as well. Table~\ref{tab:ablation_losses} summarizes the active terms in each configuration. Overall, the variants are consistently but modestly worse than the full model, by 0.5--0.9 percentage points in the average score and up to 3.8 percentage points in precision. This suggests that FACM, MMWR and RAWD act complementarily. Because the margins are small and the paired tests reach significance only for the removal of FACM (Table~\ref{tab:ablation}), we report this as a trend rather than as evidence that each module is independently essential.
\end{justify}

\begin{table}[!ht]
\centering
\small
\setlength{\tabcolsep}{6pt}
\renewcommand{\arraystretch}{1.05}
\captionsetup{justification=raggedright, singlelinecheck=false}
\caption{\bf Active loss terms in the ablation variants.}
\label{tab:ablation_losses}
\begin{tabular*}{\linewidth}{@{\extracolsep{\fill}}l c c c c@{}}
\toprule
\textbf{Configuration} & \textbf{$\mathcal{L}_{\mathrm{main}}$} & \textbf{$\mathcal{L}_{\mathrm{uni}}$} & \textbf{$\mathcal{L}_{\mathrm{dist}}$} & \textbf{$\mathcal{L}_{\mathrm{ent}}$} \\
\midrule
Full                    & \checkmark & \checkmark & \checkmark & \checkmark \\
w/o FACM                & \checkmark & \checkmark & \checkmark & \checkmark \\
w/o RAWD                & \checkmark & \checkmark & --         & \checkmark \\
w/o FACM+RAWD           & \checkmark & --         & --         & \checkmark \\
w/o MMWR+RAWD           & \checkmark & --         & --         & --         \\
\bottomrule
\end{tabular*}
\end{table}

\begin{table}[!ht]
\begin{adjustwidth}{-1.9cm}{0cm}
\captionsetup{justification=raggedright, singlelinecheck=false}
\caption{\bf Ablation study results on Task~1 of Dataset~1.}
\label{tab:ablation}
\small
\setlength{\tabcolsep}{1.5pt}
\renewcommand{\arraystretch}{1.2}
\begin{tabular*}{\linewidth}{@{\extracolsep{\fill}}l c c c c c c c c c c@{}}
\toprule
\textbf{Config.} & \textbf{ACC} & \textbf{AUPR} & \textbf{AUC} & \textbf{F1} & \textbf{Pre} & \textbf{Rec} & \textbf{Avg} & \textbf{W/T/L} & \textbf{$p$} & \textbf{F-rank} \\
\midrule
w/o FACM           & 0.9429 & 0.9735 & 0.9991 & 0.8974 & 0.9352 & 0.8758 & 0.9373 & 6/0/0 & 0.03125 & 3.17 \\
w/o RAWD           & 0.9387 & 0.9776 & 0.9989 & 0.8950 & 0.9243 & 0.8751 & 0.9349 & 5/0/1 & 0.09375 & 2.25 \\
\shortstack[l]{w/o MMWR+RAWD} & 0.9399 & \textbf{0.9783} & 0.9990 & 0.8955 & 0.9196 & 0.8793 & 0.9353 & 5/0/1 & 0.15625 & 3.17 \\
\shortstack[l]{w/o FACM+RAWD} & 0.9377 & 0.9767 & 0.9989 & 0.8928 & 0.9127 & \textbf{0.8825} & 0.9335 & 4/0/2 & 0.21875 & 2.08 \\
Full               & \textbf{0.9431} & 0.9751 & \textbf{0.9992} & \textbf{0.9048} & \textbf{0.9503} & 0.8824 & \textbf{0.9425} & --- & --- & \textbf{4.33} \\
\bottomrule
\end{tabular*}
\end{adjustwidth}
\end{table}

\subsection*{Parameter Sensitivity Analysis}
\begin{figure}[!ht]
\begin{adjustwidth}{-5.5cm}{0cm}
\centering
\includegraphics[height=4.014in]{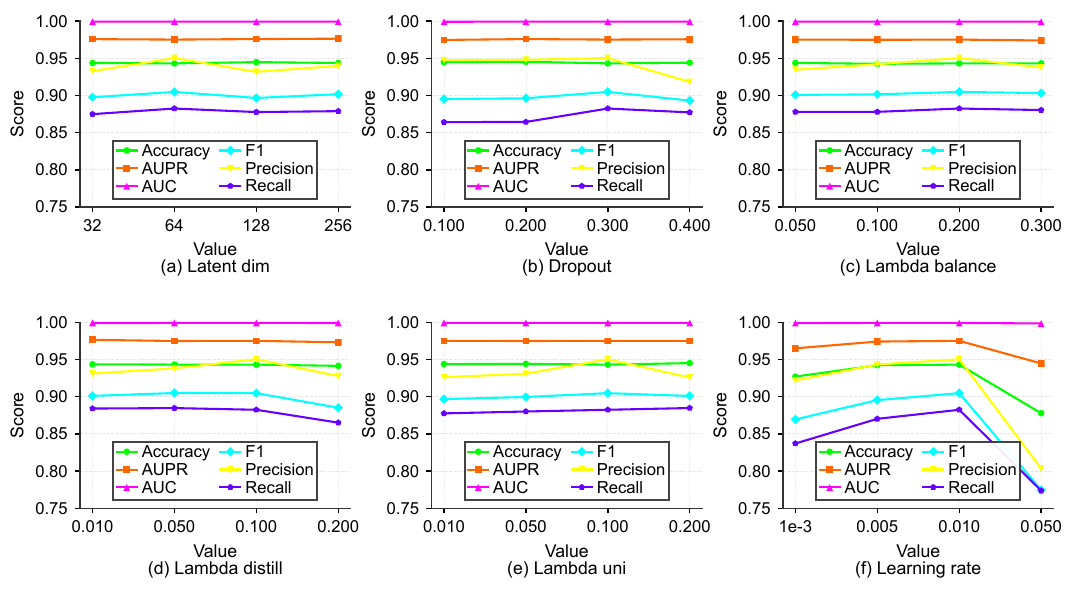}
\captionsetup{width=1.32\textwidth}
\caption{\bf Predictive performance under six hyperparameter settings on Task 1.}
\label{fig:hyperparams}
\end{adjustwidth}
\end{figure}

\begin{justify}

We vary six hyperparameters: the bottleneck dimensionality of FACM (Latent dim), the dropout rate, the routing entropy coefficient (Lambda balance), the distillation coefficient (Lambda distill), the unimodal auxiliary coefficient (Lambda uni) and the learning rate. Considering that Task 1 (known drug DDI prediction) has the largest data scale and the highest experimental stability, and that its performance trends can reliably reflect the sensitivity of the model to hyperparameter settings, we performed hyperparameter tuning on Task~1. The selected bottleneck dimension, dropout rate and loss coefficients are used in all three tasks, and the remaining training hyperparameters of Tasks~2 and 3 follow the released configuration of MKG-FENN for those settings (see the Experimental Setup).

Fig~\ref{fig:hyperparams} gives the accuracy under each configuration, which changes little within this range.
When the learning rate increases to 0.05, the performance drops notably because overly large parameter update steps cause optimization instability and prevent the model from converging to a stable optimum.
For most settings the variation in accuracy stays within 0.003, so the model is stable under these choices. We therefore set the bottleneck dimensionality to 64, the dropout rate to 0.3, the routing entropy coefficient to 0.2, the distillation coefficient to 0.1, the unimodal auxiliary coefficient to 0.1 and the learning rate to $1\times10^{-2}$.

\end{justify}

\section*{Discussion}

\subsection*{Module Contributions and Cross-Task Generalization}

\begin{justify}

We design AARME-GNN around three problems of multimodal KG-based DDI prediction: noisy and unevenly scaled unimodal features, fusion that does not adapt to individual drug pairs, and the loss of modality-specific discriminative information during integration. The results and the ablation show what each module contributes.

Across the three prediction tasks, adaptive modality routing and distillation help most when the input representations are less reliable. This is the situation with unseen drugs, whose modality-specific KGs may be noisy or incomplete. The dynamic weighting mechanism allows the model to rely more heavily on modalities that carry transferable information for unseen entities, thereby partially mitigating the cold-start problem. Identifying every true DDI event among entirely unseen drug pairs nevertheless remains difficult, as the lower recall in the complete cold-start setting shows.

The ablation study provides insight into the distinct roles of each module. FACM improves the reliability of positive predictions: by reducing within-modality noise and re-weighting channel responses in each modality, it removes variation that would otherwise produce false positives. RAWD plays a complementary role by preserving modality-specific discriminative cues that directly inform correct classifications. The variant without MMWR and RAWD ranks instances competitively but classifies them less precisely: static fusion with calibrated features orders the drug pairs well, yet without sample-adaptive routing and distillation the decision boundary is less sharp. Conversely, jointly removing calibration and distillation shifts the model toward a higher-recall, lower-precision regime, indicating that these two components work together to sharpen the decision boundary. The three modules therefore reinforce one another: FACM produces cleaner per-modality features, which sharpen MMWR's routing decisions, which in turn provide the weights for RAWD's distillation targets. The margins are small and the paired tests are significant only for the removal of FACM, so we read this pattern as a consistent trend across metrics rather than as a demonstration that each module is essential.

The architecture also offers a template for multimodal fusion under heterogeneity. Modality-specific encoders preserve the inductive bias of each data source, and a lightweight calibration module reduces within-modality noise without modifying the original encoders. Sample-adaptive routing with distillation then jointly balances modality contributions while preventing information collapse. This design philosophy may extend to other biomedical multimodal learning tasks, such as drug--target interaction prediction and polypharmacy side-effect modeling, where feature heterogeneity and sample-specific modality relevance similarly arise. AARME-GNN can screen DDI events at scale from multi-source drug information. Its predictions are candidates for further drug safety investigation, not substitutes for pharmacological experiments, clinical validation or pharmacovigilance analysis.

\end{justify}

\subsection*{Limitations and Future Work}

\begin{justify}

Several limitations remain. First, although this study covers two public DDI benchmarks, generalization to further datasets with substantially different interaction distributions, ideally proprietary or clinical data, remains to be established. Second, the relationship between the modality weights generated by MMWR and specific DDI mechanisms has not been systematically analyzed, leaving room for improved interpretability. Third, predicting interactions between entirely unseen drugs remains difficult, as indicated by the performance gap in Task~3. Fourth, the current formulation assumes that DDI events are symmetric, that is, $(d_i, d_j)$ and $(d_j, d_i)$ carry the same label, whereas some pharmacological interactions may be directional, for example when one drug primarily affects the metabolism of another. Extending the framework to directional prediction, once direction-annotated datasets exist, is a natural next step.

\end{justify}

\section*{Conclusion}
\begin{justify}

In this study we proposed AARME-GNN for multi-class DDI event prediction. By integrating FACM, MMWR and RAWD, the model calibrates unimodal representations before fusion, adaptively fuses drug pair-level features, and preserves modality-specific information. Experiments on two public DDI benchmarks under three prediction settings showed competitive performance, with consistent advantages in precision and AUC, and the ablation study supported the complementary roles of the three modules. AARME-GNN integrates multimodal knowledge graphs for DDI event prediction. Validation on datasets with different interaction distributions, a broader evaluation of the cold-start settings and better interpretability remain open.

\end{justify}

\section*{Supporting information}

\paragraph*{S1 File.}
\label{S1_File}
\justifying\setlength{\emergencystretch}{2em}%
\textbf{AARME-GNN: Code and datasets.} The compressed archive \mbox{\texttt{S1\_AARME-GNN.zip}} contains the complete source code and both benchmarks used in this study in a single package, \texttt{AARME-GNN\_Code\_and\_Datasets}. The package includes Dataset~1, with 572 drugs, 65 event types and 37{,}264 drug pairs, and Dataset~2, with 1{,}258 drugs, 100 event types and 323{,}539 drug pairs, which are selected at run time with a \texttt{--dataset} switch. It also includes a step-by-step usage guide (\texttt{README.md}) that documents the environment, the run commands and the output files for both benchmarks. The archive is organized into two directories: \texttt{code/} contains the Python implementation of AARME-GNN, and \texttt{dataset1/} and \texttt{dataset2/} contain the four-part knowledge graph datasets of the two benchmarks (Drug–Chemical Entity, Drug–Substructure, Drug–Drug, and Molecular Structure). Running the scripts writes the output files of each prediction setting to a \texttt{result/} directory. (ZIP)
\paragraph*{S1 Table.}
\label{S1_Table}
\justifying\setlength{\emergencystretch}{2em}%
\textbf{Details of the four-part knowledge graph datasets.} This table presents the complete statistics of the four heterogeneous KGs of both benchmarks used in this study, Dataset~1 with 572 drugs and Dataset~2 with 1{,}258 drugs, including the number of drugs, entities, relation types, and triplets for each subgraph. The dataset is integrated from multiple public resources and follows the same construction procedure as in MKG-FENN. (XLSX)
\paragraph*{S2 Table.}
\label{S2_Table}
\justifying\setlength{\emergencystretch}{2em}%
\textbf{Complete experimental results of parameter sensitivity analysis.} This table reports the full set of performance metrics (Accuracy, Micro AUPR, Micro AUC, Macro F1, Macro Precision, and Macro Recall) under varying configurations of six key hyperparameters: latent dimension, dropout rate, routing entropy regularization coefficient (lambda balance), distillation loss coefficient (lambda distill), unimodal auxiliary loss coefficient (lambda uni), and learning rate. All results are evaluated on Task 1 (known drug DDI prediction). (XLSX)
\paragraph*{S3 Table.}
\label{S3_Table}
\justifying\setlength{\emergencystretch}{2em}%
\textbf{Auxiliary-head input: sensitivity analysis.} This table reports the performance of AARME-GNN when the unimodal auxiliary heads are fed by either the raw modality representation specified by Eq.~(11) or the FACM-calibrated representation, for all three prediction tasks on both datasets. The differences are small and non-systematic, and the raw representation is adopted in this work. (PDF)
\section*{Data Availability Statement}
\begin{justify}
All relevant data are within the manuscript and its Supporting Information files. The Dataset~1 benchmark, a four-part knowledge graph, is released with MKG-FENN at \url{https://github.com/wudi1989/MKG-FENN}. The Dataset~2 benchmark (1{,}258 drugs, 100 DDI event types, 323{,}539 drug pairs) is the public large-scale dataset distributed with MDDI-SCL and MDF-SA-DDI at \url{https://github.com/ShenggengLin/MDF-SA-DDI}. The four knowledge-graph parts derived from it by our preprocessing pipeline, together with both benchmarks used in this work, are provided in the Supporting Information (S1 File). The authors confirm that all data underlying the findings are fully available without restriction.
\end{justify}

\section*{Code Availability Statement}
\begin{justify}
All author-generated code directly supporting this study, including data preprocessing, model implementation, training and evaluation for the three prediction tasks, is publicly available at \url{https://github.com/yjy528-06/AARME-GNN.git}. The repository contains the code of this work together with the Dataset~1 benchmark. The Dataset~2 code and data and a step-by-step usage guide are provided in the Supporting Information (S1 File, \mbox{\texttt{S1\_AARME-GNN.zip}}). The code for baseline models is not included in this repository and should be obtained from their respective original publications as cited in the manuscript.
\end{justify}

\section*{Funding}
\begin{justify}
This work was supported by the Shihezi University 2026 Undergraduate Research Training Program (SRP2026304 to JY). The funders participated in study design, data collection and analysis, decision to publish, and preparation of the manuscript.
\end{justify}

\section*{Competing interests}
\begin{justify}
The authors have declared that no competing interests exist.
\end{justify}

\bibliographystyle{plos2015}
\bibliography{cas-refs}

@article{nicholson2024prevalence,
  author = {Nicholson, K and Liu, W and Fitzpatrick, D and Hardacre, KA and Roberts, S and Salerno, J and Stranges, S and Fortin, M and Mangin, D},
  title = {Prevalence of multimorbidity and polypharmacy among adults and older adults: a systematic review},
  journal = {Lancet Healthy Longev.},
  volume = {5},
  number = {4},
  pages = {e287-e296},
  year = {2024},
  doi = {10.1016/S2666-7568(24)00007-2}
}

@article{guthrie2015rising,
  author = {Guthrie, B and Makubate, B and Hernandez-Santiago, V and Dreischulte, T},
  title = {The rising tide of polypharmacy and drug-drug interactions: population database analysis 1995--2010},
  journal = {BMC Med.},
  volume = {13},
  number = {1},
  pages = {74},
  year = {2015},
  doi = {10.1186/s12916-015-0322-7}
}

@article{mallet2007challenge,
  author = {Mallet, L and Spinewine, A and Huang, A},
  title = {The challenge of managing drug interactions in elderly people},
  journal = {Lancet.},
  volume = {370},
  number = {9582},
  pages = {185-191},
  year = {2007},
  doi = {10.1016/S0140-6736(07)61092-7}
}

@article{juurlink2003drug,
  author = {Juurlink, DN and Mamdani, M and Kopp, A and Laupacis, A and Redelmeier, DA},
  title = {Drug-drug interactions among elderly patients hospitalized for drug toxicity},
  journal = {JAMA.},
  volume = {289},
  number = {13},
  pages = {1652-1658},
  year = {2003},
  doi = {10.1001/jama.289.13.1652}
}

@article{becker2007hospitalisations,
  author = {Becker, ML and Kallewaard, M and Caspers, PWJ and Visser, LE and Leufkens, HGM and Stricker, BHC},
  title = {Hospitalisations and emergency department visits due to drug-drug interactions: a literature review},
  journal = {Pharmacoepidemiol Drug Saf.},
  volume = {16},
  number = {6},
  pages = {641-651},
  year = {2007},
  doi = {10.1002/pds.1351}
}

@article{hughes2024drug,
  author = {Hughes, JE and Moriarty, F and Bennett, KE and Cahir, C},
  title = {Drug-drug interactions and the risk of adverse drug reaction-related hospital admissions in the older population},
  journal = {Br J Clin Pharmacol.},
  volume = {90},
  number = {4},
  pages = {959-975},
  year = {2024},
  doi = {10.1111/bcp.15970}
}

@article{cosgrave2025hospital,
  author = {Cosgrave, N and Frydenlund, J and Beirne, F and Lee, S and Faez, I and Cahir, C and Williams, D},
  title = {Hospital admissions due to adverse drug reactions and adverse drug events in older adults: a systematic review},
  journal = {Age Ageing.},
  volume = {54},
  number = {8},
  pages = {afaf231},
  year = {2025},
  doi = {10.1093/ageing/afaf231}
}

@article{lin2023comprehensive,
  author = {Lin, X and Dai, L and Zhou, Y and Yu, ZG and Zhang, W and Shi, JY and Cao, DS and Zeng, L and Chen, H and Song, B and et al.},
  title = {Comprehensive evaluation of deep and graph learning on drug-drug interactions prediction},
  journal = {Brief Bioinform.},
  volume = {24},
  number = {4},
  pages = {bbad235},
  year = {2023},
  doi = {10.1093/bib/bbad235}
}

@article{ryu2018deep,
  author = {Ryu, JY and Kim, HU and Lee, SY},
  title = {Deep learning improves prediction of drug-drug and drug-food interactions},
  journal = {Proc Natl Acad Sci U S A.},
  volume = {115},
  number = {18},
  pages = {E4304-E4311},
  year = {2018},
  doi = {10.1073/pnas.1803294115}
}

@article{deng2020multimodal,
  author = {Deng, Y and Xu, X and Qiu, Y and Xia, J and Zhang, W and Liu, S},
  title = {A multimodal deep learning framework for predicting drug-drug interaction events},
  journal = {Bioinformatics.},
  volume = {36},
  number = {15},
  pages = {4316-4322},
  year = {2020},
  doi = {10.1093/bioinformatics/btaa501}
}

@article{zitnik2018modeling,
  author = {Zitnik, M and Agrawal, M and Leskovec, J},
  title = {Modeling polypharmacy side effects with graph convolutional networks},
  journal = {Bioinformatics.},
  volume = {34},
  number = {13},
  pages = {i457-i466},
  year = {2018},
  doi = {10.1093/bioinformatics/bty294}
}

@inproceedings{lin2020kgnn,
  author = {Lin, X and Quan, Z and Wang, ZJ and Ma, T and Zeng, X},
  title = {KGNN: Knowledge graph neural network for drug-drug interaction prediction},
  booktitle = {Proceedings of the Twenty-Ninth International Joint Conference on Artificial Intelligence (IJCAI)},
  year = {2020},
  pages = {2739-2745},
  doi = {10.24963/ijcai.2020/380}
}

@article{yu2021sumgnn,
  author = {Yu, Y and Huang, K and Zhang, C and Glass, LM and Sun, J and Xiao, C},
  title = {SumGNN: multi-typed drug interaction prediction via efficient knowledge graph summarization},
  journal = {Bioinformatics.},
  volume = {37},
  number = {18},
  pages = {2988-2995},
  year = {2021},
  doi = {10.1093/bioinformatics/btab207}
}

@inproceedings{su2024dual,
  author = {Su, X and Hu, P and You, ZH and Yu, PS and Hu, L},
  title = {Dual-channel learning framework for drug-drug interaction prediction via relation-aware heterogeneous graph transformer},
  booktitle = {Proceedings of the AAAI Conference on Artificial Intelligence},
  year = {2024},
  pages = {249-256},
  doi = {10.1609/aaai.v38i1.27777}
}

@inproceedings{wu2024mkg,
  author = {Wu, D and Sun, W and He, Y and Chen, Z and Luo, X},
  title = {MKG-FENN: A multimodal knowledge graph fused end-to-end neural network for accurate drug-drug interaction prediction},
  booktitle = {Proceedings of the AAAI Conference on Artificial Intelligence},
  year = {2024},
  pages = {10216-10224},
  doi = {10.1609/aaai.v38i9.28887}
}

@article{zhong2024learning,
  author = {Zhong, Y and Li, G and Yang, J and Zheng, H and Yu, Y and Zhang, J and Luo, H and Wang, B and Weng, Z},
  title = {Learning motif-based graphs for drug-drug interaction prediction via local-global self-attention},
  journal = {Nat Mach Intell.},
  volume = {6},
  number = {9},
  pages = {1094-1105},
  year = {2024},
  doi = {10.1038/s42256-024-00888-6}
}

@article{li2024multimodal,
  author = {Li, S and Tang, H},
  title = {Multimodal alignment and fusion: a survey},
  journal = {Int J Comput Vis.},
  volume = {134},
  number = {3},
  pages = {103},
  year = {2026},
  doi = {10.1007/s11263-025-02667-1}
}

@article{planamente2024relative,
  author = {Planamente, M and Plizzari, C and Peirone, SA and Caputo, B and Bottino, A},
  title = {Relative norm alignment for tackling domain shift in deep multi-modal classification},
  journal = {Int J Comput Vis.},
  volume = {132},
  number = {7},
  pages = {2618-2638},
  year = {2024},
  doi = {10.1007/s11263-024-01998-9}
}

@article{shen2025benchmarking,
  author = {Shen, Z and Zhou, M and Zhang, Y and Yao, Q},
  title = {Benchmarking drug-drug interaction prediction methods: a perspective of distribution changes},
  journal = {Bioinformatics.},
  volume = {41},
  number = {11},
  pages = {btaf569},
  year = {2025},
  doi = {10.1093/bioinformatics/btaf569}
}

@article{wu2025medical,
  author = {Wu, J and Wang, Z and Hong, M and Ji, W and Fu, H and Xu, Y and Xu, M and Jin, Y},
  title = {Medical SAM adapter: Adapting segment anything model for medical image segmentation},
  journal = {Med Image Anal.},
  volume = {102},
  pages = {103547},
  year = {2025},
  doi = {10.1016/j.media.2025.103547}
}

@article{he2025adaptfrcnet,
  author = {He, A and Wu, Y and Wang, Z and Li, T and Fu, H},
  title = {AdaptFRCNet: Semi-supervised adaptation of pre-trained model with frequency and region consistency for medical image segmentation},
  journal = {Med Image Anal.},
  volume = {103},
  pages = {103626},
  year = {2025},
  doi = {10.1016/j.media.2025.103626}
}

@inproceedings{hu2018squeeze,
  author = {Hu, J and Shen, L and Sun, G},
  title = {Squeeze-and-excitation networks},
  booktitle = {Proceedings of the IEEE Conference on Computer Vision and Pattern Recognition},
  year = {2018},
  pages = {7132-7141},
  doi = {10.1109/CVPR.2018.00745}
}

@article{vilar2014similarity,
  author = {Vilar, S and Uriarte, E and Santana, L and Lorberbaum, T and Hripcsak, G and Friedman, C and Tatonetti, NP},
  title = {Similarity-based modeling in large-scale prediction of drug-drug interactions},
  journal = {Nat Protoc.},
  volume = {9},
  number = {9},
  pages = {2147-2163},
  year = {2014},
  doi = {10.1038/nprot.2014.151}
}

@article{chen2021muffin,
  author = {Chen, Y and Ma, T and Yang, X and Wang, J and Song, B and Zeng, X},
  title = {MUFFIN: multi-scale feature fusion for drug-drug interaction prediction},
  journal = {Bioinformatics.},
  volume = {37},
  number = {17},
  pages = {2651-2658},
  year = {2021},
  doi = {10.1093/bioinformatics/btab169}
}

@inproceedings{lyu2021mdnn,
  author = {Lyu, T and Gao, J and Tian, L and Li, Z and Zhang, P and Zhang, J},
  title = {MDNN: A Multimodal Deep Neural Network for Predicting Drug-Drug Interaction Events},
  booktitle = {Proceedings of the Thirtieth International Joint Conference on Artificial Intelligence (IJCAI)},
  year = {2021},
  pages = {3536-3542},
  doi = {10.24963/ijcai.2021/487}
}

@article{lin2022mdf,
  author = {Lin, S and Wang, Y and Zhang, L and Chu, Y and Liu, Y and Fang, Y and Jiang, M and Wang, Q and Zhao, B and Xiong, Y and et al.},
  title = {MDF-SA-DDI: predicting drug-drug interaction events based on multi-source drug fusion, multi-source feature fusion and transformer self-attention mechanism},
  journal = {Brief Bioinform.},
  volume = {23},
  number = {1},
  pages = {bbab421},
  year = {2022},
  doi = {10.1093/bib/bbab421}
}

@article{deng2022meta,
  author = {Deng, Y and Qiu, Y and Xu, X and Liu, S and Zhang, Z and Zhu, S and Zhang, W},
  title = {META-DDIE: predicting drug-drug interaction events with few-shot learning},
  journal = {Brief Bioinform.},
  volume = {23},
  number = {1},
  pages = {bbab514},
  year = {2022},
  doi = {10.1093/bib/bbab514}
}

@article{nyamabo2021ssi,
  author = {Nyamabo, AK and Yu, H and Shi, JY},
  title = {SSI-DDI: substructure-substructure interactions for drug-drug interaction prediction},
  journal = {Brief Bioinform.},
  volume = {22},
  number = {6},
  pages = {bbab133},
  year = {2021},
  doi = {10.1093/bib/bbab133}
}

@article{zhang2022molormer,
  author = {Zhang, X and Wang, G and Meng, X and Wang, S and Zhang, Y and Rodriguez-Paton, A and Wang, J and Wang, X},
  title = {Molormer: a lightweight self-attention-based method focused on spatial structure of molecular graph for drug-drug interactions prediction},
  journal = {Brief Bioinform.},
  volume = {23},
  number = {5},
  pages = {bbac296},
  year = {2022},
  doi = {10.1093/bib/bbac296}
}

@article{su2022attention,
  author = {Su, X and Hu, L and You, Z and Hu, P and Zhao, B},
  title = {Attention-based knowledge graph representation learning for predicting drug-drug interactions},
  journal = {Brief Bioinform.},
  volume = {23},
  number = {3},
  pages = {bbac140},
  year = {2022},
  doi = {10.1093/bib/bbac140}
}

@article{hong2022lagat,
  author = {Hong, Y and Luo, P and Jin, S and Liu, X},
  title = {LaGAT: link-aware graph attention network for drug-drug interaction prediction},
  journal = {Bioinformatics.},
  volume = {38},
  number = {24},
  pages = {5406-5412},
  year = {2022},
  doi = {10.1093/bioinformatics/btac682}
}

@inproceedings{xiong2023multi,
  author = {Xiong, Z and Liu, S and Huang, F and Wang, Z and Liu, X and Zhang, Z and Zhang, W},
  title = {Multi-relational contrastive learning graph neural network for drug-drug interaction event prediction},
  booktitle = {Proceedings of the AAAI Conference on Artificial Intelligence},
  year = {2023},
  pages = {5339-5347},
  doi = {10.1609/aaai.v37i4.25665}
}

@article{li2026ka,
  author = {Li, Y and Liu, JM and Li, YC and You, HR and You, ZH and Yuan, Y and Mi, CG},
  title = {KA-DDI: A knowledge-adaptive contrastive learning framework for drug-drug interaction prediction},
  journal = {Expert Syst Appl.},
  volume = {308},
  pages = {131076},
  year = {2026},
  doi = {10.1016/j.eswa.2025.131076}
}

@inproceedings{fang2025emoe,
  author = {Fang, Y and Huang, W and Wan, G and Su, K and Ye, M},
  title = {EMOE: Modality-specific enhanced dynamic emotion experts},
  booktitle = {Proceedings of the IEEE/CVF Conference on Computer Vision and Pattern Recognition},
  year = {2025},
  pages = {14314-14324},
  doi = {10.1109/cvpr52734.2025.01335}
}

@article{baranwal2020deep,
  author = {Baranwal, M and Magner, A and Elvati, P and Saldinger, J and Violi, A and Hero, AO},
  title = {A deep learning architecture for metabolic pathway prediction},
  journal = {Bioinformatics.},
  volume = {36},
  number = {8},
  pages = {2547-2553},
  year = {2020},
  doi = {10.1093/bioinformatics/btz954}
}

@article{lin2022mddi,
  author = {Lin, S and Chen, W and Chen, G and Zhou, S and Wei, DQ and Xiong, Y},
  title = {MDDI-SCL: predicting multi-type drug-drug interactions via supervised contrastive learning},
  journal = {J Cheminform.},
  volume = {14},
  number = {1},
  pages = {81},
  year = {2022},
  doi = {10.1186/s13321-022-00659-8}
}

@article{lee2019novel,
  author = {Lee, G and Park, C and Ahn, J},
  title = {Novel deep learning model for more accurate prediction of drug-drug interaction effects},
  journal = {BMC Bioinformatics.},
  volume = {20},
  number = {1},
  pages = {415},
  year = {2019},
  doi = {10.1186/s12859-019-3013-0}
}

@inproceedings{kipf2016semi,
  author = {Kipf, TN and Welling, M},
  title = {Semi-supervised classification with graph convolutional networks},
  booktitle = {International Conference on Learning Representations (ICLR)},
  year = {2017},
  doi = {10.48550/arXiv.1609.02907}
}

@article{li2025kgmaeddi,
  author = {Li, Y and You, ZH and Yang, Y and Mi, CG},
  title = {KGMAEDDI: Knowledge Graph and Molecular-Graph Masked Autoencoder for Drug-Drug Interaction Prediction},
  journal = {IEEE J Biomed Health Inform.},
  volume = {PP},
  pages = {1-11},
  year = {2025},
  doi = {10.1109/JBHI.2025.3610870}
}

@article{ma2023dual,
  author = {Ma, Mei and Lei, Xiujuan},
  title = {A dual graph neural network for drug-drug interactions prediction based on molecular structure and interactions},
  journal = {PLoS Comput Biol.},
  volume = {19},
  number = {1},
  pages = {e1010812},
  year = {2023},
  doi = {10.1371/journal.pcbi.1010812}
}

@article{zhu2026heterogeneous,
  author = {Zhu, H and Wang, J and Hua, Z and Wang, C and Zhang, Z and Yu, T and Ge, L},
  title = {Heterogeneous biological graph convolutional network for drug-target interaction prediction},
  journal = {PLoS One.},
  volume = {21},
  number = {5},
  pages = {e0348895},
  year = {2026},
  doi = {10.1371/journal.pone.0348895}
}

@inproceedings{hinton2015distilling,
  title     = {Distilling the knowledge in a neural network},
  author    = {Hinton, Geoffrey and Vinyals, Oriol and Dean, Jeff},
  booktitle = {NIPS Deep Learning and Representation Learning Workshop},
  year      = {2015}
}

@article{gou2021knowledge,
  author  = {Gou, Jianping and Yu, Baosheng and Maybank, Stephen J. and Tao, Dacheng},
  title   = {Knowledge distillation: a survey},
  journal = {Int J Comput Vis.},
  year    = {2021},
  volume  = {129},
  number  = {6},
  pages   = {1789--1819},
  doi     = {10.1007/s11263-021-01453-z}
}

@inproceedings{furlanello2018born,
  title        = {Born again neural networks},
  author       = {Furlanello, Tommaso and Lipton, Zachary and Tschannen, Michael and Itti, Laurent and Anandkumar, Anima},
  booktitle    = {Proceedings of the 35th International Conference on Machine Learning},
  series       = {Proceedings of Machine Learning Research},
  volume       = {80},
  pages        = {1607--1616},
  year         = {2018},
  organization = {PMLR}
}

@inproceedings{zhang2019be,
  title     = {Be your own teacher: improve the performance of convolutional neural networks via self distillation},
  author    = {Zhang, Linfeng and Song, Jiebo and Gao, Anni and Chen, Jingwei and Bao, Chenglong and Ma, Kaisheng},
  booktitle = {Proceedings of the IEEE/CVF International Conference on Computer Vision},
  year      = {2019},
  pages     = {3712--3721},
  doi       = {10.1109/ICCV.2019.00381}
}

\end{document}